\documentstyle[12pt,emlines,bezier,epsf,eqsecnum,aps]{revtex}
\tightenlines
\begin{document}
\draft
\title{Operator Product Expansion on a Fractal:\\
The Short Chain Expansion for Polymer Networks}
\author{Christian von Ferber}
\address{Fachbereich Physik, Universit\"at Essen, D-45117 Essen\\
Tel.: +49 201 183-2987  Fax: +49 201 183-2120\\
email: cerbo@theo-phys.uni-essen.de}
\maketitle
\begin{abstract}
We prove to all orders of renormalized perturbative polymer field theory
the existence of a short chain expansion applying to polymer
solutions of long and short chains.
For a general polymer network with long and short chains we show factorization
of its partition sum by a short chain factor and a long chain factor in the
short chain limit.
This corresponds to an expansion for short distance along the fractal
perimeter of the polymer chains connecting the network vertices and is related
to a large mass expansion of field theory.

The scaling of the second virial coefficient for bimodal solutions is 
explained.
Our method also applies to the correlations of the multifractal measure of
harmonic diffusion onto an absorbing polymer. 
We give a result for expanding these correlations for short
distance along the fractal carrier of the measure.
\end{abstract}
\section{Introductory Part}\label{I}
\label{Kap01}
\subsection{Introduction}\label{Ia}
Operator product expansions (OPEs) have served as a crucial tool of analysis
ever since their general concept was introduced by Wilson \cite{Wilson69}.
They have been elaborated for many systems which allow for a description
in terms of local field operators. In their original form they describe
correlations of observables at short distance or large momenta in terms of
another local observable. More generally, the basic hypothesis is that
products of local operators $\phi_{\rm A}, \phi_{\rm B}$ of a theory
at short distance may be expanded in terms of other local operators
$\phi_{\rm C}$
\begin{equation}
\phi_{\rm A}(x)\phi_{\rm B}(x+r) = \sum_{\rm C} a_{\rm ABC}(r)
\phi_{\rm C}(x) \label{1a1}
\end{equation}
with coefficients $ a_{\rm ABC} $ absorbing all divergencies in the limit
$r \to 0$.
If the scaling properties of the operators are described by scaling
dimensions $y_A$ (i.e., rescaling all lengths by $\lambda$ and operators
by $\lambda^{y_A}\phi_A(\lambda x)$ the theory remains invariant)
then the behavior of the leading term $a_{ABC}(r)$ for $r\to 0$
is described by a power law with an obvious scaling relation for the
exponent.
\begin{equation}
\phi_A(x)\phi_B(x+r)\sim a_{ABC}(r)\phi_C(x)
,\quad a_{ABC}(r) \sim r^{\Theta_{ABC}}
,\quad \Theta_{ABC} = y_A + y_B - y_C.
\end{equation}
 The short distance expansion (SDE) in eq. (\ref{1a1}) defines
an algebra of operators of the theory which in fact is essential for the
analysis of conformal field theories, for a review see \cite{Cardy87}.

How is this concept applied in polymer theory?
The theory of polymers in solution, modelled
as selfrepelling walks, defines mean values
of observables by averages over all spatial configurations of the polymer
chains.
For selfrepelling walks of $N$ steps on a $d$ dimensional lattice
typical results are power laws for the end - end distance
$R\sim N^{\nu}$ and the number of configurations
${\cal Z}\sim e^{\mu N}N^{\gamma - 1}$ with  a lattice dependent fugacity
$e^\mu$. The universal exponents $\nu$ and $\gamma$ may be derived as $n=0$ 
limits of the corresponding exponents of $O(n)$ symmetric $\phi^4$ field
theory \cite{DeGennes72}.
We may generalize the theory to `$f$ - stars', i.e., configurations of $f$
polymer chains constrained to have one common endpoint.
For polymer stars of $f$ chains of equal
length $N$, the number of configurations scales like
${\cal Z}_f\sim e^{\mu Nf}N^{\gamma_f-1}$, defining a family of exponents
$\gamma_f$ with $\gamma_2=\gamma_1=\gamma$ 
\cite{Myake83,Duplantier86,Ohno88}.
The exponents $\gamma_f$ again may be derived as $n=0$ limits of
corresponding exponents in $O(n)$ symmetric $\phi^4$ field theory
\cite{Schaefer92,Wallace75}.
If we now consider the probability $P(r)$ of finding the cores of two
$f_1$- and $f_2$- stars at distance $r\ll R$ we find a power law
\begin{equation}\label{I2}
P_{f_1f_2}(r)\sim r^{\Theta} \quad,\quad
\nu\Theta=(\gamma_{f_1}-1)+(\gamma_{f_2}-1) - (\gamma_{f_1+f_2}-1)
\quad .
\end{equation}
This law corresponds to the leading term in the operator product expansion
of field theory.
The cores of the $f$- stars are represented by composite operators
$\phi_f \sim (\phi)^f$ with appropriate symmetry and scaling dimensions
related to $\gamma_f$, such that the above law can be derived from
the relation
\begin{equation}\label{I3}
\phi_{f_1}(x)\phi_{f_2}(x+r) \sim r^{\Theta}\phi_{f_1+f_2}(x)\quad.
\end{equation}
It should be noted, that the relation (\ref{I2}) includes as special
cases the relations for the contact exponents $\Theta$, introduced by
des Cloizeaux \cite{desCloizeaux80} to describe the behavior of the
probabilities $P_{12}$, $P_{22}$ to find a chain end and a monomer
or any two monomers at distance $r$.

\begin{figure}
\epsfbox{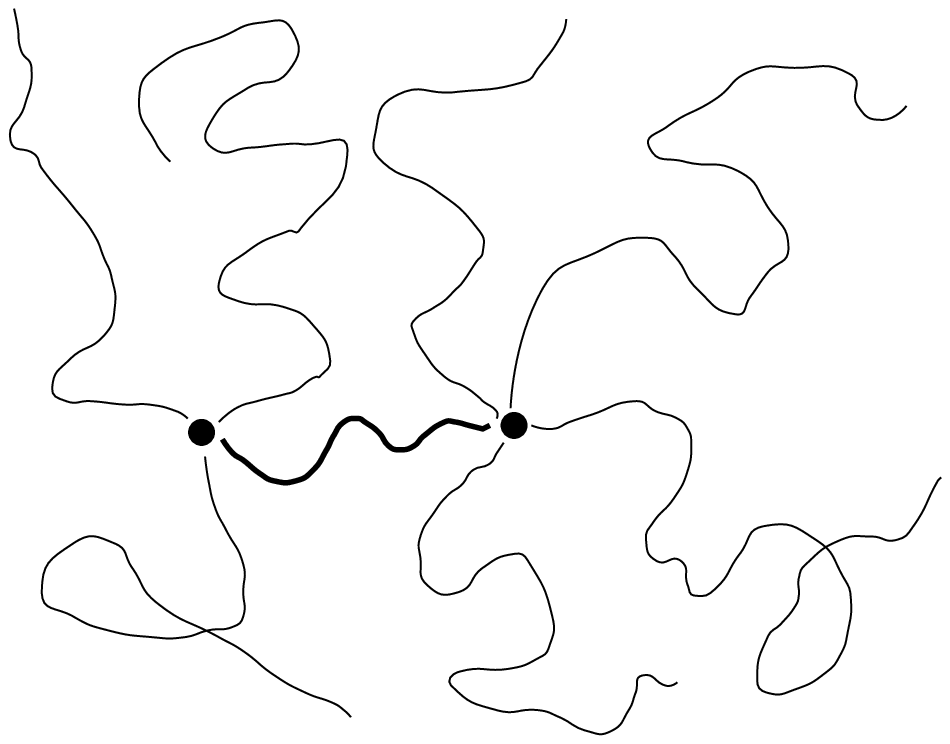} 
\caption{\label{figId}
Network ${\cal G}_0$ of two stars connected by a short `bridge'
$\cal A$. When $\cal A$ is contracted to a vertex a star 
${\cal G}_0/{\cal A}$ results.}
\end{figure}

\begin{figure}
\epsfbox{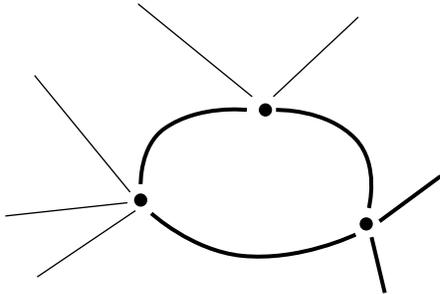}
\caption{\label{figIa}
Graph of a network ${\cal G}_0$ which will contract to a star, when the 
subnetwork $\cal A$ of short chains (bold) is contracted to a vertex.}
\end{figure}

The short distance expansion describes the behavior of polymer stars
when their cores are at short distance in space.
Here, in contrast,
we want to discuss the properties of polymer stars which are
linked by short chains between their cores.
We will expand the partition sum of such networks for the chains between
the cores being much shorter than the outer chains of the stars.
In general,
linking the chain endpoints of a set of polymers, one may even construct
networks of any topology. For a network ${\cal G}_0$ made from
$F$ chains of fixed length $N$, the number of configurations
${\cal Z}_{{\cal G}_0}$ again follows a power law
\cite{Duplantier86,Duplantier89,Schaefer92}
\begin{equation}\label{I4}
{\cal Z}_{{\cal G}_0}(N)\sim e^{\mu NF} N^{\gamma_{{\cal G}_0} -1}
\quad,\quad
\gamma_{{\cal G}_0} -1 =
\nu d L_{{\cal G}_0} + \sum_{f\geq 1} n_f
[(\gamma_f - 1) - f/2(\gamma - 1)]
\quad.
\end{equation}
Here $L_{{\cal G}_0}$ is the number of loops in ${{\cal G}_0}$ and the $n_f$
give the numbers of vertices with $f$ legs.
All configurations are counted, unregarding the question of possible knots
in $d=3$. Thus for our purpose any network is characterized by a graph.
Consider now a network ${\cal G}_0$ in which one or more chains are much 
shorter than the others. Call the subnetwork of short chains $\cal A$.
Shrinking $\cal A$ to one vertex we receive a new network
${\cal G}_0/{\cal A}$; we will assume that it is a star.
Examples are given in figs. \ref{figId} and \ref{figIa}.
In view of the above `contact' laws, we will expect the number of 
configurations
of ${\cal G}_0$ to decrease with the length of the short chains.
With numbers of monomers $N_1$ of short chains and $N_2$ of long chains,
we propose for $N_1\ll N_2$
\begin{equation}\label{I5}
{\cal Z}_{{\cal G}_0}(N_1,N_2) \sim N_1^{\nu \Theta}
{\cal Z}_{{\cal G}_0/{\cal A}}(N_2)
\quad,\quad
\nu\Theta = \gamma_{{\cal G}_0} - \gamma_{{\cal G}_0/{\cal A}}
\quad.
\end{equation}
Having in mind the description of network vertices by composite operators
$\phi_f$, this is now an OPE for short `distance' along the fractal
perimeter of the short chains; so it looks like an OPE on a fractal.

In this paper we present a proof for
the existence of this non trivial expansion in terms of field theory.
We will show that the leading term of the expansion in $N_1/N_2\ll 1$
has the form (\ref{I5}) for all orders of renormalized perturbation
theory.

\begin{figure}
\epsfbox{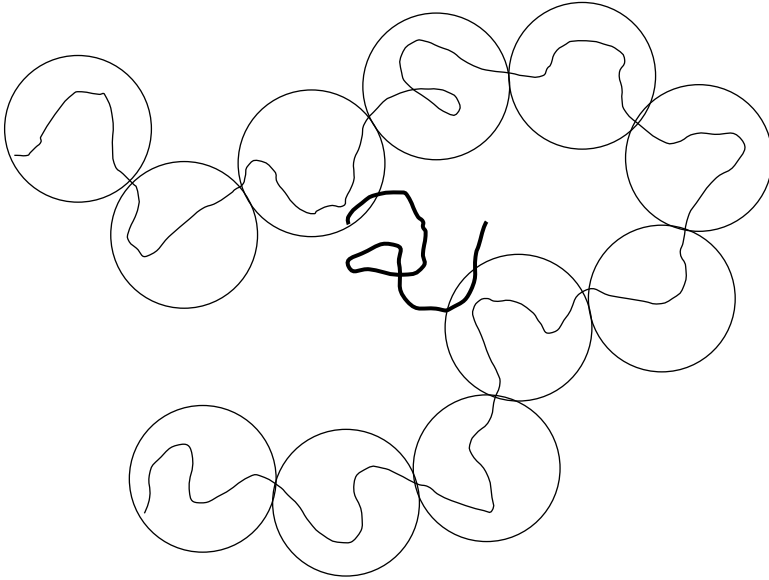}
\caption{\label{figIb}
Interaction of a long polymer chain and a short polymer chain
in solution. In the blob picture the short chain interacts only with
one `blob' of its own size at the time. Here we take into account that
any part of the long chain can come near the short chain.}
\end{figure}

Clearly, this short chain expansion is also relevant for the behavior
of `bimodal' polymer solutions, containing two polymer species of
very different lengths \cite{Krueger89,Witten82}.
To give an example, which is also accessible experimentally,
 we consider the second virial coefficient of the
osmotic pressure.
The osmotic pressure $\Pi$ can be developed in the
concentrations $c_1$ and $c_2$ of the short and long chains.
The virial coefficients of this expansion may then be
determined again as universal scaling functions.
With Boltzmann constant $k_{\rm B}$ and temperature $T$ we write
\begin{equation}\label{I6}
\frac{\Pi}{k_B T}= c_1 + c_2 + A^{(2)}_{11}c^2_1 +A^{(2)}_{22}
c^2 _2 +A^{(2)}_{12}c_1 c_2 + \dots\quad.
\end{equation}
The second virial coefficient $A^{(2)}_{12}$ describes the interaction of the
long chains with the short chains (see fig \ref{figIb}).
The second virial coefficients $A_{ij}^{(2)}$ have the dimension
of a volume and, assuming scaling invariance of (\ref{I6}),
$A_{ii}^{(2)}\sim R_i^d$ where $R_i$ is the radius of the $i$th polymer
coil.
Following the argument of \cite{Witten82}, a scaling law for the mixed
coefficient can be derived when the number of monomers $N_2$ of the
long chains is much larger than the number of monomers $N_1$ of the
short chains, $N_2\gg N_1$.
Then the short chain `sees' the long chain as $\frac{N_2}{N_1}$ `blobs'
of its own size (see fig. \ref{figIb}), each blob corresponding to a chain of
length $N_2$.
Following this argument we would expect
\begin{equation}\label{I7}
 A_{12}(N_1 N_2) \sim \frac{N_2}{N_1}R_1^d
\end{equation}
The blob picture gives only a very incomplete description of the actual
situation. The short chain can interact with parts of the long chain
which are very far apart along the chain. Therefore, a thorough analysis
is needed, including the interaction at high orders, to show that these
contributions will cancel in the limit considered.
This analysis we perform in general for networks with chains of
differing lengths. We show that the scaling law (\ref{I7}) follows from these
general consideration as a special case (see \ref{IVb}).

The OPE as a successful concept is often also applied by analogy in spite of
the fact that a description in terms of a local field (theory) has not been
shown or even might be doubtful. The spatial correlations in multifractal 
distributions as they occur e.g. in turbulence theory are analyzed
by scaling relations analogous to those derived from an OPE.
Attempts have been made to define the appropriate ``field" operators
\cite{Eyink93}.
Let us explain the idea for a general multifractal measure $\mu$ 
\cite{Cates87a}.
A measure $\mu(x)$ defined on a set of linear
size $R$ is called multifractal if the moments of $\mu$
scale with independent powers $d_n$ for small cutoff $a$
\cite{Halsey86}:
\begin{equation}\label{I8}
 \langle \mu(x)^n \rangle \sim (R/a) ^{d_n}
\end{equation}
The spatial correlations of different moments of the measure are generally
believed to have a short distance behavior analogous to an OPE:
\begin{equation} \label{I9}
  \langle \mu(x)^m\mu(x+r)^n \rangle \sim r^{\Theta_{mn}}
   \langle \mu(x)^{m+n} \rangle
\mbox{\qquad with \qquad}
 \Theta_{mn} = d_m+d_n - d_{m+n} .
\end{equation}
The scaling relations for the exponents $\Theta_{mn}$ have been derived
on the basis of blob arguments \cite{Cates87a}.
Our analysis of the SCE for networks directly applies to the special case
of the multifractal measure defined from harmonic diffusion onto an
absorbing polymer as introduced in \cite{Cates87}.

The paper is organized as follows.
The investigations are performed on a continuous model defined in
the next section.
Star polymers  as well as their relation to the composite operators are 
discussed. The last section \ref{Ic} of this chapter
introduces the problem of the short chain expansion using a simple example.
In the chapters \ref{II} and \ref{III},
we give the proof for the existence of the
first coefficient of this expansion. The problem of graph algebra and
summation is explained in chapter \ref{II}.
The analytical properties of these sums are discussed in chapter \ref{III}.
Consequences of the results of this work we discuss in chapter
\ref{IV}. On the one hand, it explains the scaling of the
second virial coefficient in solutions
of long and short polymer chains.
On the other hand, it sheds new light on the
problem of multifractal correlations in the case of the
flux of harmonic diffusion onto a polymer.
\subsection{Model, notation, and the trivial one loop case}\label{Ib}
The polymers in solution we describe by the continuous chain model,
as introduced by Edwards \cite{Edwards65,Edwards66}. In this model,
the configuration of a polymer chain $a$ is given by a path $r^a(s)$
in $d$ dimensional space $I\!\!R^d$, parametrized by a variable
$0 \le s \le S_a$.
The statistical weight of the configuration of a set $a=1,\ldots,F$ of
$F$ chains is determined by the Hamiltonian $\cal H$
divided by the product of Boltzmann constant $k_{\rm B}$ and temperature
$T$,
\begin{equation}\label{1b1}
\frac{1}{k_{\rm B}T}{\cal H} =
\frac{1}{2} \sum_{a=1}^{F}\int_0^{S_a}{\rm d}s
\left( \frac{{\rm d}r^a(s)}{{\rm d}s} \right)^2
+\frac{u}{2} \sum_{a,b=1}^{F}
\int{\rm d}^d r \rho_a(r) \rho_b(r)
\quad,
\end{equation}
\begin{equation}\label{1b2}
\mbox{with densities} \hspace{10mm}
\rho_a(r) = \int_0^{S_a}{\rm d}s \delta^d(r-r^a(s))
\quad.
\end{equation}
Obviously, the variables $s_a, S_a$ have the dimension of a surface.
$S_a$ is called the Gau\ss{}ian surface of chain $a$; here we will
often refer to $S_a$ as the `length' of the chain.
The first term of the Hamiltonian $\cal H$ describes the connectivity
along the chain by elastic forces. The second term models the interaction
between points on the chains. It is of short range with strength $u$ and by
(\ref{1b2}) even point like. Of course this model by no means gives a
realistic picture of the microscopic structure of any real polymer.
By renormalization group analysis rather it is shown that it is a most
simple representative of the universality class of all more realistic models
for specific systems of polymers in solution.

The partition sum is calculated as a functional integral over all possible
configurations of the polymer system normalized to unity for vanishing
interaction $u$. As a simple network let us consider a star with an
$F$-leg vertex connecting $F$ polymer chains at one end. Its partition
sum ${\cal Z}_F$ has to respect the constraint that all chains have one
common endpoint $r^a(0)$:
\begin{equation}\label{1b3}
{\cal Z}_F\{S_a\}=\frac{1}{{\cal Z}_F^0} \int{\rm D}[r^a(s)]
\exp\{-\frac{1}{k_{\rm B}T} {\cal H}\}
\prod_{a=2}^{F} \delta^d(r^a(0)-r^1(0))
\quad.
\end{equation}
The normalization ${\cal Z}_F^0$ is given by the same functional integral
evaluated for vanishing coupling $u=0$. It is a Gau\ss{}ian integral
and describes $F$ random walks with mean square end-end distances
$R^2_a=S_a d$, $(a=1,\ldots,F)$, all starting at the same point.
For nonvanishing coupling $u$, a regularization is needed to define the
integral (\ref{1b3}). Introducing a cutoff $s_0$, such that all simultaneous
integrals in $s_a,s'_a$ are cut off for $|s_a-s'_a|\ge s_0$, the theory may
be evaluated perturbatively. In terms of this cutoff we may view the polymer
chain as a self repelling walk with $N_a=S_a/s_0$ microscopic steps.
Bare perturbation theory strongly depends on this microscopic cutoff.
To extract the universal content in the limit of long chains,
it is necessary to eliminate this dependence
by transformation to a renormalized theory.
To this end several equivalent approaches exist. Rather than performing
direct renormalization \cite{Duplantier86,desCloizeaux75,desCloizeaux90}
of the polymer variables, we chose
here to make use of language and formalism of local field theory,
which is more apt to our intentions. For
regularization and renormalization we use a dimensional scheme in the
notation of \cite{Bergere81}. The mapping to field theoretic formalism
is performed by a Laplace transform in the Gau\ss{}ian surfaces $S_a$
to conjugate chemical potentials, `mass variables' $\mu_a$,
\begin{equation}\label{1b4}
\tilde{\cal Z}_F \{\mu_a\} = \int\prod_b{\rm d}S_b e^{-\mu_b S_b}
{\cal Z}_F\{S_a\}
\quad.
\end{equation}
To construct a local field theory, we transform from the functional integral
over all configurations $r^a(s)$ to an integral over a local field.
Then the density $\rho_a(r)$ is described in terms
of the square $\phi^2(r)$ of the local field.
In the corresponding Lagrangean field theory, the transformed
partition sum is given as a zero-momentum correlator. In real space
with volume $\Omega$ it reads
\begin{equation}\label{1b5}
\tilde{\cal Z}_F\{\mu_a\} = \frac{1}{\Omega} \int\prod_{b=0}^F {\rm d}^d r_b
\left\langle\prod_{a=1}^F \phi_a(r_0) \prod_{a=1}^F\phi_a(r_a)
\right\rangle^{{\cal L}_n} |_{n\to 0} \;.
\end{equation}
The correlator is evaluated with the $O(n)$ symmetric Lagrangean
${\cal L}_n$ of $F$ fields, each with $n$ components,
\begin{equation}\label{1b6}
{\cal L}_n\{\phi_a,\mu_a\} = \frac{1}{2}\sum_{b=1}^F \int{\rm d}^d r
\left( \mu_b\phi_b^2(r) + \frac{1}{2}(\nabla\phi_b(r))^2 \right)
+ \frac{u}{8} \sum_{bb'}^F \int{\rm d}^d r \phi^2_b(r)\phi^2_{b'}(r)
\quad.
\end{equation}
Loosely speaking, the Laplace transform maps the exponential
propagator of the polymer theory in (\ref{1b1}) to the propagator
of the Lagrangean theory, whereas the mapping $\rho_a \to \phi^2_a$
explains the correspondence of the interaction terms in $\cal H$
and ${\cal L}_n$. For the properties of this transform see the
book of des~Cloizeaux and Jannink \cite{desCloizeaux90}
and for our purposes also \cite{Schaefer91}.
The formal limit $n=0$ in (\ref{1b5}) eliminates bubble graph contributions
in the perturbation expansion which have combinatorial factors of powers
of $n$, thus producing the `polymer limit' of $O(n)$ symmetric $\phi^4$ field
theory. The local composite operator $\prod_{a=1}^F \phi_a(r_0) $
describes in this limit the $F$ functional core of the polymer star.

We illustrate these definitions with an example which is also suitable
to introduce our short chain expansion.
Let us evaluate the perturbation expansion of ${\cal Z}_F$ to first loop order.
The corresponding integrals may be represented by graphs as in fig.
\ref{figIe}. For the full Feynman rules see appendix \ref{A1}.
In Gau\ss{}ian normalization and dimensional regularization it reads
\begin{eqnarray}
{\cal Z}_F\{S_a\} &=& 1 - u \sum_{a=1}^F I_1(S_a)
- u \sum_{a<b}^F \dot{I}_1^{ab} + {\cal O}(u^2) \quad, \\
\mbox{with \hspace{3em}}
I_1(S) &=& \int_0^S {\rm d}s (S-s) s^{-d/2}\quad,
\end{eqnarray}
where we have omitted numerical factors occurring with the same power as $u$.
 Obviously, the graph for
$\dot{I}_1^{ab}$ is constructed from $I_1$ by inserting the vertex on the
inner line of a chain of length $S_a+S_b$. This is reflected in the
relation
\begin{equation}
 \dot{I}_1^{ab} = I_1(S_a+S_b) - I_1(S_a) - I_1(S_b)\quad,
\end{equation}
In dimensional regularization and by analytic continuation we can
evaluate the integrals for $d=4-\epsilon$ with $\epsilon>0$.
We want to study the limit in which one arm of the star, say $S_F$, is
much shorter than the others $a=1,\ldots,F-1$.
Clearly, this limit here concerns only $\dot I_{aF}$, and from the
explicit results we trivially find to leading $\epsilon$ order 
\begin{eqnarray}
 I_1(S) &=& S^{\frac{\epsilon}{2}} (\frac{2}{\epsilon} + 1) \quad,\\
\dot I_1^{aF}(S_a,S_F) &=&
 \left[(S_a +S_F)^{\epsilon/2} - S_a^{\epsilon/2} - S_F^{\epsilon/2}
 \right]
(\frac{2}{\epsilon} + 1)
\stackrel{S_F/S_a\to 0}{\longrightarrow} - I_1(S_F)
\quad.
\end{eqnarray}
In the limit $S_F/S_a\to 0$, $a=1,\ldots,F-1$, the partition sum for the
$F$-star factorizes into the $F-1$-star partition sum and a factor only
depending on the short chain length $S_F$:
\begin{eqnarray}
{\cal Z}_F\{S_a\} &=& \left[ 1 - u I_1(S_F) - u \sum_{a=1}^F \dot I_1^{aF} 
\right]
{\cal Z}_{F-1}\{S_a\} + {\cal O}(u^2) \nonumber\\
&\stackrel{S_F/S_a\to 0}{\longrightarrow}&
\left[ 1 + u (F-2) I_1(S_F)  \right] {\cal Z}_{F-1}\{S_a\} + {\cal O}(u^2)
\quad.
\end{eqnarray}
Of course, here this factorization is just a consequence of the fact
that the limit exists for the one loop amplitude that contributes.
Already in two loop order the integrals of single contributions do not
converge in this limit. We will see that appropriate cancellations
occur which make the factorization well defined again.
\subsection{A non trivial two loop example}\label{Ic}
\label{S1c}
Let us discuss the problem for the first nontrivial case in some
detail. It will already contain the idea of the general proof in principle.
Again we consider the $F$-star with chain $F$ much shorter than
the others.
To reduce the considerations to the essential point we discuss the zero
momentum vertex function $\Gamma_F\{S_a\}$.
Here we simply define $\Gamma_F\{S_a\}$ to be the sum of contributions
of $F$-vertex polymer graphs; i.e., one line irreducible (1li) polymer graphs
containing the $F$-vertex.
To first loop order there is only one such graph, labeled $\dot I_1^{ab}$
in fig.\ref{figIe}.
It is constructed from the graph labeled $I_1$ by inserting the vertex into
the inner line.
All two loop contributions involving two arms are constructed in the same
way from the graphs labeled $I_2$, $I_3$, inserting the vertex into all inner
lines.This gives rise to integrals $\dot I_2^{ab}$, $\dot I_3^{ab}$ by the
formula
\begin{equation}
  \dot I_k^{ab} = I_k(S_a+S_b) - I_k(S_a) - I_k(S_b) \quad,
  \hspace{4em} k=1,2,3.
\end{equation}
There is only one  contribution involving three arms. It is is labeled
$I_4^{abc}$ in fig.\ref{figIe}.
To second loop order we thus write the vertex function $\Gamma_F$ as
\begin{equation}
\Gamma_F\{S_a\} = 1 - u \sum_{a<b}^F \dot I_1^{ab}
 + u^2 \sum_{a<b}^F (\dot I_2^{ab} + \dot I_3^{ab})
 + u^2 \sum_{a\ne b\ne c}^F I_4^{abc}
 + u^2 \sum_{a<b\ne c<d}^F \dot I_1^{ab}\dot I_1^{cd} +{\cal O}(u^3)
.
\end{equation}
We want to show that in the short chain limit, where
 $S_F/S_{a\le F-1}\to 0$,
the vertex function $\Gamma_F$ factorizes into $\Gamma_{F-1}$ and a factor
${\cal C}^*_F$ depending on the short chain length $S_F$ only:
\begin{equation}
\Gamma_F\{S_{a'}\}_{{a'}\le F} \stackrel{S_F/S_a\to 0}{\longrightarrow}
{\cal C}_F^*(S_F) \Gamma_{F-1}\{S_a\}_{a\le F-1}
\quad.
\end{equation}
To construct this factor we formally divide 
${\cal C}_F= \Gamma_{F} / \Gamma_{F-1}$ 
\begin{equation}\label{1.22}
{\cal C}_F = 1 + \sum_{a=1}^{F-1} ( -u \dot I_1^{aF}
 + u^2 \dot I_2^{aF} + u^2 \dot I_3^{aF})
 + u^2 \sum_{a\ne b}^{F-1} (I_4^{Fab}+I_4^{aFb}+I_4^{abF}-
\dot I_1^{ab}\dot I_1^{aF}) + {\cal O}(u^3)
\quad.
\end{equation}
The two loop integrals $I_k$ have the form
$I_k(S)\sim S^\epsilon a_k(\epsilon)$, see \cite{Duplantier89}, so that
the short chain limit exists for $\dot I_k^{aF}$ by construction, as
explicitly shown for the one loop integral in the last paragraph.
As we will see below, the three arm contribution $I_4^{abF}$
does not converge in this limit, but produces a singularity
which is cancelled by the subtracted term in (\ref{1.22}).
Let us take a closer look at this integral,
\begin{equation}\label{1.23}
I_4^{abc} = \int_0^\infty {\rm d}s_4\cdots{\rm d}s_1
\left[(s_1+s_2)(s_3+s_4)+s_1s_2\right]^{-d/2}
\Theta(S_a-s_4-s_1)\Theta(S_b-s_3)\Theta(S_c-s_2)
\quad,
\end{equation}
with the Heaviside $\Theta(\ldots)$ - function, and the term in
square brackets being the Symanzik polynomial of the graph.
To further analyse the problem, we partition the integration volume into
Hepp sectors $s_{\sigma(4)}>s_{\sigma(3)}>s_{\sigma(2)}>s_{\sigma(1)}$,
labeled by permutations $\sigma \in {\wp}_4$ of the four lines.
We first restrict ourselves to the sector $\sigma_0:s_4>\cdots>s_1$
and perform a convenient substitution
\begin{eqnarray}\label{1.24}
\lefteqn{
I_4^{abc}(\sigma_0) = 2^4 \int_0^\infty {\rm d}\beta_4
\int_0^1 {\rm d}\beta_3{\rm d}\beta_2{\rm d}\beta_1
\beta_4^{-\omega_4-1}\beta_3^{-\omega_3-1}
\beta_2^{-\omega_2-1}\beta_1^{-\omega_1-1}
}&&
\\ \nonumber &&
\left[(1+\beta_1^2)(1+\beta_3^2)+\beta_3^2\beta_2^2\beta_1^2\right]^{-d/2}
\Theta(S_a-\beta_4^2-\beta_4^2\beta_3^2\beta_2^2\beta_1^2)
\Theta(S_b-\beta_4^2\beta_3^2)\Theta(S_c-\beta_4^2\beta_3^2\beta_2^2)
\quad.
\end{eqnarray}
The substitution rule may be read off from the arguments of the $\Theta$ -
functions. The exponents $\omega_i$ result from the Jakobian of the
transformation, and from factorizing out common factors in the
Symanzik polynomial. They give the degree of divergence of the subgraphs
$R_i=\{1,\ldots,i\}$ of lines $1,\ldots,i$. Simple power counting gives
\begin{equation}
\omega_4 = 2d -8, \hspace{3em}
\omega_3 = d -6, \hspace{3em}
\omega_2 = d -4, \hspace{3em}
\omega_1 = -2\quad.
\end{equation}
In $d<4 $ all $\omega_i$ are negative and the integral converges.

We want to study the short chain limit of ${\cal C}_F$ in
(\ref{1.22}). Let us investigate the situation that $c=F$ is the short
chain in (\ref{1.24}). The $\beta_4$ integral may be performed in the
limit $S_a,S_b\to \infty$ and gives a $\Gamma$- function:
\begin{eqnarray}\label{1.26}
I_4^{abF}(\sigma_0) = 2^3 \Gamma(-\frac{\omega_4}{2})
S_F^{-\frac{\omega_4}{2}}
\int_0^1 {\rm d}\beta_3{\rm d}\beta_2{\rm d}\beta_1
\beta_3^{\omega_4-\omega_3-1}
\beta_2^{\omega_4-\omega_2-1}\beta_1^{-\omega_1-1}\nonumber\\
\left[(1+\beta_1^2)(1+\beta_3^2)+\beta_3^2\beta_2^2\beta_1^2\right]^{-d/2}
\end{eqnarray}
Let us define `important' subgraphs as those subgraphs which
fully contain the short chain $F$. This property reflects the fact
that their $\beta$ - parameters will appear in the remaining
$\Theta(S_F - \beta_4^2\beta_3^2\beta_2^2)$ - function, and will
be decorated with an additional $\omega_4$ exponent after the
$\beta_4$ integration is performed. Here $R_2$ and $R_3$ are
important subgraphs.
From (\ref{1.24}),(\ref{1.26}) we deduce that for the integral
$I_4^{abF}(\sigma_0)$ the limit exists, if \cite{Bergere76}
\begin{itemize}
\item[(i)] all subgraphs are convergent,
$\omega_1,\omega_2,\omega_3,\omega_4<0$;
\item[(ii)] `important' subgraphs $R_2,R_3$ are more convergent than the
total graph, $\omega_2,\omega_3 < \omega_4$ .
\end{itemize}
This result is generalized in the next section and in appendix \ref{B}.
Obviously, here, condition (ii) is violated, as $\omega_2>\omega_4$
for dimensions $d<4$. The essential point is now, to see that including
the subtraction generated by formally inverting $\Gamma_{F-1}$  in
(\ref{1.22}),
we receive a convergent sum of integrals also in the limit
$S_a,S_b\to\infty$.
Using the same substitutions as above, now for the product of integrands of
the product $\dot I_1^{ab}\dot I_1^{bF}$, we have
\begin{eqnarray}\label{1.27}
(I_4^{abF}-\dot I_1^{ab}\dot I_1^{bF})(\sigma_0) =
2^3 \Gamma(-\frac{\omega_4}{2})
S_F^{-\frac{\omega_4}{2}}
\int_0^1 {\rm d}\beta_3{\rm d}\beta_2{\rm d}\beta_1
\beta_3^{\omega_4-\omega_3-1}
\beta_2^{\omega_4-\omega_2-1}\beta_1^{-\omega_1-1}
\nonumber\\ \{
\left[(1+\beta_1^2)(1+\beta_3^2)+\beta_3^2\beta_2^2\beta_1^2\right]^{-d/2}
-\left[(1+\beta_1^2)(1+\beta_3^2)\right]^{-d/2}
\}\quad.
\end{eqnarray}
For the subtracted integrand of (\ref{1.27}), the singularities for
$\beta_3\to 0$, $\beta_2\to 0$ cancel.
Let us briefly explain how this result may be
found for all Hepp sectors of the subtracted integral
\begin{eqnarray}\label{1.28}
I_4^{abF}-\dot I_1^{ab}\dot I_1^{bF} &=&
\int {\rm d}s_i \Theta(S_F-s_2) J^{\rm ps}(s_1,\ldots,s_4)
\quad,\\ 
J^{\rm ps}(s_1,\ldots,s_4)&=&
\left[ (s_1+s_2)(s_3+s_4)+s_1s_2\right]^{-d/2}
- (s_1+s_2)^{-d/2}(s_3+s_4)^{-d/2}
\quad.
\end{eqnarray}
To this end, we generalize the notion of degree of divergence, to include
integrands of the form $J^{\rm ps}$. For a subgraph given by a collection
of lines $R_\ell=\{1,\ldots,\ell\}$, the degree of divergence
of the function $J$ is $\omega_{R_\ell}(J)$ if
\begin{equation}\label{1-37}
\rho^{2\ell+\omega_{R_\ell}} J(\rho^2s_1,\ldots,\rho^2s_\ell,
s_{\ell+1},\ldots,s_n) = c_0 + c_1 \rho + c_2 \rho^2 + \cdots
\hspace{2em}\mbox{ for some } c_0\neq 0
\quad.
\end{equation}
By permutation this is generalized to any subgraph
$R\subseteq\{1,\ldots,n\}={\cal G}$ of the total graph $\cal G$.
For our example we find
\begin{equation}
\omega_{R_1}(J^{\rm ps}) = -4, \hspace{2em}
\omega_{R_2}(J^{\rm ps}) = d-6, \hspace{2em}
\omega_{R_3}(J^{\rm ps}) = d-6, \hspace{2em}
\omega_{\cal G}(J^{\rm ps}) = 2d-8 \quad.
\end{equation}
Conditions (i) and (ii) now apply by the same argument as above to yield
a convergent integral on Hepp sector $\sigma_0$.
It is easy to verify that condition (i) holds for all subgraphs and indeed
(ii) for all important subgraphs $R$,
\[
 \{2\}\subseteq R \subset {\cal G} ,
  \omega_R(J^{\rm ps}) < \omega_{\cal G}(J^{\rm ps})
\quad.
\]
Hence we can take over the argument for any permutation $\sigma$ of
$(1,\ldots,4)$, to show convergence on all Hepp sectors.
The same method applies to the remaining integrals
$I_4^{Fab}$ and $I_4^{aFb}$, which converge without subtraction in the limit
$S_a,S_b\to \infty$.

To show that the short chain limit exists for all orders of (\ref{1.22}),
we will have to prove that the appropriate subtractions take place
at all orders.
Let us write this subtraction as a partial sum (ps) of amplitudes in the
following more general notation
\begin{equation}
I_{\cal G}^{\rm ps} = I_4^{abF} - \dot{I}_1^{ab} \dot{I}_1^{aF}
= I_{\cal G} - I_{{\cal G}/{g}}I_{g}
\quad.
\end{equation}
Here $g$ is the important subgraph $g=R_2$ of $\cal G$, for which
${\cal G}/{g}$ is a star graph contribution to $\Gamma_{F-1}$
of the reduced `network' of long chains.
This is illustrated in fig.\ref{figIf}.
The generalization of this notion of partial sums of amplitudes
and the cancellation of divergences will be subject of the next chapters.
We will then show that this scheme commutes with renormalization and the
Laplace transform.
\section{Partial Summation}
\label{Kap08}\label{II}

In our investigation in two loop order, we have
already seen that, for certain single contributions to
${\cal C}_{F}$,
the short chain limit does not exist. So we are seeking a
generalization to the partial sum

\[ {I}_{\cal G}^{ps} = {I}_{\cal G} -{I}_{{\cal G}/g} {I}_g \quad,
\]
where ${\cal G}$ is a contributing graph and ${\cal G}$ has a subgraph $g$,
which contains all lines of short chains. If $g$ has additional subgraphs
with this property, we will also need additional subtractions to receive a
finite limit, and so forth.  In this way, we are naturally led to consider 
nests of subgraphs, these are ordered families of subgraphs in which every 
element is subgraph of the preceding element.\\
Following this idea, let us consider the algebra of subtractions occurring for 
a general network ${\cal G}_0$. Let ${\cal G}_0$ be a
network with long and short chains, the short chains forming a subnetwork
${\cal A}$ such that pulling together the subnetwork ${\cal A}$ to a vertex a 
star graph ${\cal G}_0/{\cal A}$ results, see fig.\ref{figIa}.
We can formally factorize the vertex function of the network, 
$\Gamma_{{\cal G}_0}$, into the vertex function of the star, 
$\Gamma_{{\cal G}_0/{\cal A}}$, and a factor ${\cal C}_{{\cal G}_0,{\cal A}}$.
We will show here that we can represent
${\cal C}_{{\cal G}_0,{\cal A}}$ in terms of partial sums each of which is
constructed from the nests of subgraphs of a single graph contributing to
${\cal C}_{{\cal G}_0,{\cal A}}$.

The recursive structure of the nests, and therefore also of the partial sums,
will allow us to prove the existence of the limit to all orders of perturbation
theory in chapter \ref{III}.
Let us introduce some notation first. The set of non trivial polymergraphs,
contributing to the vertex function of the network,
${\Gamma}_{{\cal G}_{0}}$, we will denote by $[{\cal G}_{0}]$,
the set of nontrivial polymergraphs contributing to 
the vertex function of the star, ${\Gamma}_{{\cal G}_0 /{\cal A}}$, we will 
denote by $[{\cal G}_0 / {\cal A} ]$.

With the above definitions, we can write the perturbation series for the
vertex part of the partition sum as a sum over contributions ${I}_{\cal G}$ of 
graphs ${\cal G}$:
\begin{eqnarray}
  {\Gamma}_{{\cal G}_{0}} && =  {I}_{{\cal G}_{0}}
 + \sum_{{\cal G} \in [{\cal G}_{0}]}
u^{\cal G}
{I}_{\cal G}
 =
\sum_{{\cal G} \in [{\cal G}_{0}]^0}
u^{\cal G}
{I}_{\cal G}
\quad, \\
  {\Gamma}_{{\cal G}_{0}/{\cal A}}&&
 =
{I}_{{\cal G}_{0}/{\cal A}}
 + \sum_{{\cal G} \in [{\cal G}_{0}/{\cal A}]}
u^{\cal G}
 {I}_{\cal G}
 \quad.
\end{eqnarray}
here we have written ${\cal G} \in [{\cal G}_{0}]^0 $ for ${\cal G}$ being 
either a
non-trivial graph ${\cal G} \in [{\cal G}_{0}]$, or the trivial graph ${\cal G}
= {\cal G}_0$.  We have put the factor $u^{\cal G}$, to denote the appropriate
power of the interaction parameter $u$, and any other factors occurring in the
power of the loop order of ${\cal G}$.
We expand formally
\begin{equation}\label{II3}
{\cal C}_{{\cal G}_0,{\cal A}} =
\frac{{\Gamma}_{{\cal G}_{0}}} {{\Gamma}_{{\cal G}_{0}/{\cal A}}} =
{\Gamma}_{{\cal G}_{0}} \sum_{n=0}^\infty
{(1-{\Gamma}_{{\cal G}_{0}/{\cal A}})}^n
\quad.
\end{equation}
Expanding the powers of $n$, we will now reorder the products of contributions
of graphs. Taking the normalization $ {I}_{{\cal G}_{0}/{\cal A}} = 1$, we
have
\begin{eqnarray}
{\cal C}_{{\cal G}_{0},{\cal A}}&& =
\sum_{g_{0} \in [{\cal G}_{0}]^0}
 u^{g_0}
{I}_{g_0}
 {(1 + \sum_{g \in [{\cal G}_{0}/{\cal A}]}
u^{g}
 {I}_{g})}^{-1}
\\
&& = \sum_{{g_{0} \in {[{\cal G}_{0}]^0}}}
u^{g_0}
{I}_{g_{0}}
\sum_{n=0}^\infty {(-\sum_{g \in {[{\cal G}_{0}/{\cal A}]}}
u^{g}
{I}_{g} )}^n
\\
&& =  \sum_{g_{0},g_{1}^{0},g_{2}^{0},\dots,g_{n}^{0}}
u^{g_0}
 {I}_{g_{0}}
\prod_{k=1}^n
(-
  u^{g_k^0}
{I}_{g_{k}^0})
\quad.
\label{8.6}
\end{eqnarray}
In this way, we have represented ${\cal C}_{{\cal G}_{0},{\cal A}}$, as a sum
over all families of polymer graphs $(g_{0}, g_{1}^{0},...,g_{n}^{0})$, $n\geq
0$, where the first element is the trivial or a non trivial graph $g_{0} \in
[{\cal G}_{0}]^0$, and all others are non trivial graphs
$g_{1}^{0},...,g_{n}^{0} \in [{\cal G}_{0}/{\cal A}]$.  The
summation is over all families, respecting the order of not necessarily
different elements $g_{1}^{0},...,g_{n}^{0}$.

To every family of graphs, we will now construct a nest of subgraphs of a graph
${\cal G} \in [{\cal G}_{0}]^0$.  To the family of graphs
$g_{0},g_{1}^{0},...,g_{n}^{0}$, we will construct the nest of subgraphs $g_{0}
\subset g_{1} \subset ... \subset g_{n}={\cal G}$ of the graph ${\cal G} \in
[{\cal G}_{0}]^0$.

We construct $g_{1}$ by replacing in $g_{1}^{0}$ the star vertex by the graph
$g_0$. The  path structure of the polymer chains of ${\cal G}_{0}$ shows which
exterior leg of $g_0$ corresponds to which leg of the vertex in $g_{1}^{0}$.
Next, we construct $g_2$,  replacing the star vertex in the graph $g_{2}^{0}$
by the just constructed graph $g_1$, and so forth.
In general, we construct the graphs $g_k$, $k=1...n$, replacing the star
vertex in $g_{k}^{0}$ by the before constructed graph $g_{k-1}$.
The relation of the family of $g_k^0$ graphs to the nest of graphs $g_k$ is
most easily understood from the inverse formula
\begin{equation}
g_{k}^{0} =g_{k}/ g_{k-1}\quad.
\end{equation}
Each element of the $g_k^0$ family shows what is added, going from one
level of the nest $g_{k-1}$ to the next higher level $g_{k}$.

Let us explain the procedure, using a simple example:
In fig.~\ref{figIIa}
 a family of graphs $(g_{0}, g_{1}^{0},g_{2}^{0})$ is depicted in
the first line, and the corresponding nest of subgraphs  $(g_{0}, g_{1},g_{2})$
in the second line. The short chain is depicted as a bold line.

From the above considerations we conclude:
There is a one to one correspondence of families of graphs
in the sum of (\ref{8.6}) defined by
\begin{equation}
(g_{0}, g_{1}^{0},...,g_{n}^{0}) \quad\mbox{with}\quad
 g_{0} \in [{\cal G}_{0}]^0, g_k^0 \in [{{\cal G}^{0}}/{\cal A}]
\quad,\quad k=1,\ldots,n\quad,
\end{equation}
to nests of subgraphs of a graph $\cal G$, contributing to 
$\Gamma_{{\cal G}_0}$,
with the following properties
\begin{equation}\label{8.9}
g_{0} \subset g_{1} \subset ...\subset g_{n}={\cal G}\quad\mbox{with}\quad
g_{k} \in [{\cal G}_{0}]^0, g_{k}/g_{k-1}  \in
[{\cal G}_{0}/{\cal A}],\quad k=1,\ldots,n \quad.
\end{equation}
The set of nests of subgraphs to a given graph ${\cal G} \in [{\cal G}_{0}]$,
with the property (\ref{8.9}), we denote as ${\cal N}_{\cal A}({\cal G})$.
Now we can rewrite (\ref{8.6}) as a sum over products of all nests of subgraphs
of the same graph ${\cal G}\in [{\cal G}_0]^0$:
\begin{eqnarray}
{\cal C}_{{\cal G}_{0}, {\cal A}}
&& = \sum_{{\cal G}\in[{\cal G}_0]^0}%
u^{\cal G}%
     \sum_{(g_{0}...g_{n}) \in {\cal N}_{\cal A}({\cal G})}
     \prod_{k=1}^{n} (-{I}_{ g_{k}/g_{k-1} }){I}_{g_{0}}   \\
&& = \sum_{{\cal G}\in[{\cal G}_0]^0}
     u^{\cal G}
     {I}_{\cal G}^{ps}
\end{eqnarray}
The partial sum of contributions ${I}^{ps}_{\cal G}$ belonging to the
graph ${\cal G}$ is defined here as a sum over all nests in ${\cal N}_{\cal
A}({\cal G})$.
We will now construct a recursive formulation for $I_{\cal G}^{\rm ps}$.
Let $g$ be a subgraph of $\cal G$,
contributing to ${\Gamma}_{{\cal G}_{0}}$; i.e.,
$g \in [{\cal G}_{0} ]^0$.
Let the graph $\cal G$ be transformed to a star graph when pulling $g$ 
together to a vertex, while this star graph ${\cal G}/g$
contributes to the
star partition sum ${\Gamma}_{{\cal G}_{0} / {\cal A}}$; i.e.,
${\cal G}/g \in [{\cal G}_{0} / {\cal A}]$. In the following, we will denote
these properties of $g$ by the short notation
\[
  g \prec {\cal G} :
\quad g \subset {\cal G} \mbox{ with } g \in [{\cal G}_{0} ]^0
\mbox{ and } {\cal G}/g \in [{\cal G}_{0} / {\cal A}]
\;\;
\]
and call $g$ a  `star subgraph' of $\cal G$.
For the partial sums we then find the recursive representation
\begin{equation}
{I}_{\cal G}^{\rm ps} = {I}_{\cal G} - \sum_{g \prec {\cal
G}}{I}_{{\cal G}/g} {I}_{g}^{\rm ps}
\end{equation}
Recursive expansion of this formula produces a sum of products of contributions
${I}_{g_{k}/g_{k-1}}$ for graphs $g_{k}$, which for each product form a nest,
by construction. This representation of the factor
${\cal C}_{{\cal G}_{0}, {\cal A}}$
is the essential key for a recursive proof of cancellations to
all orders of (renormalized) perturbation theory, as performed in the next
chapter (\ref{III}).
\section{The Short Chain Limit}
\label{III}
\subsection{Existence below $d=4$}
\label{IIIa}
In chapter \ref{II}, for any network ${\cal G}_0$, that contracts to
a star ${\cal G}_0/{\cal A}$ when the subnetwork of short chains
contracts to a vertex, we have generalized the factorization of the
network vertex function $\Gamma_{{\cal G}_0}$ into the star vertex
function $\Gamma_{{\cal G}_0/{\cal A}}$ and a formal factor
${\cal C}_{{\cal G}_0,{\cal A}}$.
For the contributions to this factor in perturbation theory we have
found several representations. In particular we have constructed a
recursive formula which associates to each single graph, contributing
to $\Gamma_{{\cal G}_0}$, a sum over nests of subgraphs. This generates
the products of graph amplitudes, corresponding to the expansion of the
formal ratio
${\cal C}_{{\cal G}_0,{\cal A}}=
\Gamma_{{\cal G}_0}/\Gamma_{{\cal G}_0/{\cal A}}$.
In the following we will use this recursive form of the contributions
to the factor ${\cal C}_{{\cal G}_0,{\cal A}}$, to show that the
short chain limit of the factorization of $\Gamma_{{\cal G}_0}$
exists.
Thus we will prove that indeed, in the short chain limit, the
network vertex function $\Gamma_{{\cal G}_0}$ factorizes into a
long chain part $\Gamma_{{\cal G}_0/{\cal A}}$ and a short chain
part ${\cal C}^*_{{\cal G}_0,{\cal A}}$ (the star indicates the limit).
This is interpreted as the first term of an expansion of the
network vertex function $\Gamma_{{\cal G}_0}$ for short chains in the
subnetwork $\cal A$.
To perform the short chain limit on ${\cal C}_{{\cal G}_0,{\cal A}}$
we take infinite length $S_f\to\infty$ for the long chains while
the short chains remain finite. In the following we denote this
limit by a star.
In the short chain limit all integrals over long chain variables
in $I_{\cal G}^{*\rm ps}$ have infinite upper limit.
$I_{\cal G}^{\rm ps}$ is a sum of products of integrals. In each
product all but one factor are integrals over long chain variables.
In the short chain limit only the constraints on the short chain
variables of the last factor remain. This factor corresponds to a
star subgraph $g\prec{\cal G}$ of $\cal G$, which contains all
`short lines' $a\in A_f$ on short chains $f\in {\cal A}$.
For any chain $f\in {\cal G}$, we denote the set of lines on this
chain by $A_f$.
Taking the limit, the constraints are the same for all terms of
$I_{\cal G}^{*\rm ps}$ and we can write it as one integral
\begin{equation}\label{III1}
I_{\cal G}^{*\rm ps} = \int_0^{\infty} \prod_{a\in {\cal G}}{\rm d}s_a
\prod_{f\in {\cal A}} \Theta^{(j_f)} (S_f - \sum_{a\in A_f} s_a)
\Delta_{\cal G}^{\rm ps}
\quad \mbox{with}\quad
\Delta_{\cal G}^{\rm ps} = \Delta_{\cal G}^{-d/2} -
\sum_{g\prec{\cal G}}\Delta_{{\cal G}/g}^{-d/2}\Delta_{g}^{\rm ps}
\quad,
\end{equation}
with $\Theta$ - functions $j_f=0$ for outer chains and $\delta$ - functions
$j_f=1$ for inner chains of $\cal A$, and the Symanzik polynomials
$\Delta_{\cal G}$,
$\Delta_g$ of ${\cal G}$ and its subgraphs (see appendix \ref{A1}).
The definition of partial sums $\Delta_g^{\rm ps}$ is understood by
recursion.
The degree of divergence $\omega_g(\Delta_{\cal G}^{\rm ps})$
of any subgraph $g$ of $\cal G$ is defined as in (\ref{1-37}).
Obviously, there is no problem to define $\omega_g$ simultaneously for
any family or nest of subgraphs, $\Delta_{\cal G}^{\rm ps}$
has the so called simultaneous Taylor property \cite{Bergere76}.
We want to investigate the convergence properties of the integral
$I_{\cal G}^{\rm ps *}$ near the upper critical dimension $d=4$.
First, we have to take care of propagator loops in $\cal G$,
subgraphs which are divergent above $d=2$.
As explained in appendix \ref{A3}, we use dimensional regularization
and perform the analytical continuation of the integrand by
subtracting the appropriate truncated Taylor expansions. Below
$d=4$, only lowest order subtractions occur which will also affect
the product of $\Theta$ - functions. Although one could now go on,
working with the resulting products of distributions
(they are well defined, see appendix \ref{A5}), the notion of degree of
divergence is more clear for the Laplace transformed integral.
We transform only in the remaining short chain variables  and receive
an integral with vanishing mass parameters $\mu_f$ for the long chains,
and finite $\mu_f$ for short chains.
\begin{eqnarray}
\tilde{I}_{\cal G}^{\rm ps *} &=& \int\prod_{f\in {\cal A}}
{\rm d} S_f
\prod_{f\in {\cal A}}\mu_f^{1-j_f} e^{-\mu_f S_f}
I_{\cal G}^{\rm ps *}(\{S_f)\})
\nonumber\\ \label{III2}
&=& \int\prod_{a\in {\cal G}}{\rm d}s_a \tilde{J}_{\cal G}^{\rm ps *}
\quad,\quad
\tilde{J}_{\cal G}^{\rm ps *} =
\exp(-\sum_{f\in {\cal A}}\sum_{a\in A_f} \mu_f s_a) \Delta_{\cal G}^{\rm ps}
\quad.
\end{eqnarray}

Let us note here, that although we have derived (\ref{III1}),(\ref{III2})
for vertex function contributions, i.e., one line irreducible graphs $\cal G$, 
they apply as well if $\cal G$ may be disconnected cutting one line.
Subgraphs separated from the vertex component in this way will lie on dangling
long chains, and both integrands $J^*_{\cal G}$, $\tilde J^*_{\cal G}$
factorize with respect to these subgraphs. Some elementary algebra then shows
that the partial sum will vanish \cite{Ferber93}. It follows that 
the limit ${\cal C}^*_{{\cal G}_0,{\cal A}}$ remains unchanged, if one
replaces vertex functions $\Gamma_{{\cal G}_0}$ by partition sums
${\cal Z}_{{\cal G}_0}$ in (\ref{II3}).  

For the following, we consider $\tilde{J}_{\cal G}^{\rm ps *}$ to be the
analytical continuation in $d$ for $2<d<4$ from below $d=2$. The corresponding
subtractions only affect the propagator subgraphs.
Once made convergent, these subgraphs play no role for our further
argument. To save notational overhead, we will ignore these subtractions,
and presume that all subgraphs $g\subset {\cal G}$
have negative degree of divergence with respect to
$\tilde{J}_{\cal G}^{\rm ps *}$ 
for $2<d<4$.
In appendix \ref{B} we show by generalization of the criteria found in chapter
\ref{I} that the integral converges if, with respect to
$\tilde{J}_{\cal G}^{\rm ps *}$
\begin{itemize}
\item[(i)] all subgraphs $g$ of $\cal G$ are convergent: 
 $\omega(\tilde{J}_{\cal G}^{\rm ps *}) < 0$,
\item[(ii)] all important subgraphs $R$ 
    (containing all short lines of $\cal G$)
are more convergent than the full graph $\cal G$: 
$\omega_R < \omega_{\cal G}$.
\end{itemize}
Property (i) is fulfilled for the analytical continuation in $2<d<4$.
Let us now show the property (ii) for the important subgraphs of a graph 
$\cal G$ by the inductive method.

{\em Step0: } We anchor the induction, showing (ii) for graphs $\cal G$
which do not contain any star subgraphs $g \prec {\cal G}$.
Their partial sum has only one term
$\tilde{J}_{\cal G}^{\rm ps *} = \tilde{J}_{\cal G}^{\rm *}$,
the integrand of the graph $\cal G$ for vanishing mass parameter of the
long chain variables.
Let $\cal G$ be a graph contributing to $\Gamma_{{\cal G}_0}$
which has no star subgraphs $g \prec {\cal G}$. Assume $R\subset {\cal G}$
is an important subgraph and, with respect to $\tilde{J}_{\cal G}^{\rm *}$,
is at least as divergent as $\cal G$. Let $g_0$ be the important
connected component of $R$; then
\begin{equation}\label{III3}
\omega_{g_0   }(\tilde{J}_{\cal G}^{\rm *}) \ge
\omega_{R     }(\tilde{J}_{\cal G}^{\rm *})  \ge
\omega_{\cal G}(\tilde{J}_{\cal G}^{\rm *})\quad.
\end{equation}
We show $g_0\prec{\cal G}$, invalidating this assumption.
For the degrees of divergence of $g_0$ and $\cal G$, we may ignore
the above mentioned propagator subtractions and calculate 
$\omega$, starting from
the number of loops or number of interaction vertices $v_g$ and external
legs $q_g$. From topological relations one derives (see \ref{A1}),
\begin{equation}\label{III4}
\omega_{g_0}(\tilde{J}_{\cal G}^{\rm *})
- \omega_{\cal G}(\tilde{J}_{\cal G}^{\rm *}) =
(4-d)(v_{\cal G} - v_{g_0}) + (1-\frac{d}{2})(q_{g_0}-q_{\cal G})
\quad.
\end{equation}
As $g_0$ contains all lines belonging to the short chain subnetwork
$\cal A$, it connects all vertices of the underlying network ${\cal G}_0$.
Thus $g_0$ has at least one external line on every external chain of
${\cal G}_0$, $q_{g_0}\geq q_{{\cal G}_0} = q_{\cal G}$.
The number of vertices of $g_0$ is smaller than of $\cal G$ such that
$v_{g_0}\leq v_{\cal G}$.  Then from (\ref{III3}) and (\ref{III4})
follows $q_{g_0} = q_{\cal G}$; $g_0$ has the same number of external legs
as $\cal G$. This insures that the important subgraph $g_0$ respects star
topology in the sense $g_0\prec {\cal G}$, in contradiction to the first
assumption and so, all subgraphs $R$ must respect (ii).

{\em Step1:} Our second anchor are all those graphs $\cal G$ which have exactly
one star subgraph $g_0\prec{\cal G}$. With the argument of Step0 the
connected component of any important subgraph $R$ that could violate
(ii) is $g_0$. Now the Symanzik polynomial $\Delta_{\cal G}$ for $\cal G$ may
be decomposed with respect to the subgraph $g_0$ (see \ref{A1}) such that
\begin{equation}\label{III5}
\Delta_{\cal G}^{\rm ps} = \left[ \Delta_{{\cal G}/g_0}\Delta_{g_0}
+ {\cal P}_{{\cal G},g_0} \right]^{-\frac{d}{2}}
- \Delta_{{\cal G}/g_0}^{-\frac{d}{2}}\Delta_{g_0}^{-\frac{d}{2}}
\quad.
\end{equation}
Here the polynomial ${\cal P}_{{\cal G},g_0}$ is of higher minimal
degree in the $\ell_{g_0}$ variables of $g_0$ than the polynomial
$\Delta_{g_0}$, which has homogenous degree $L_{g_0}$ with
$\omega_{g_0}(\Delta_{g_0}^{-\frac{d}{2}}) = d L_{g_0} - 2 \ell_{g_0}$.
Therefore, we find
\begin{eqnarray}\label{III6}
\omega_{g_0}(\tilde{J}_{\cal G}^{\rm ps *}) &\leq&
\omega_{g_0}(\Delta_{g_0}^{-\frac{d}{2}}) - 2
\quad. \\ \nonumber
\omega_{g_0}(\tilde{J}_{\cal G}^{\rm ps *})
-\omega_{\cal G}(\tilde{J}_{\cal G}^{\rm ps *}) &\leq&
(4-d)(v_{\cal G} - v_{g_0}) - 2 \quad.
\end{eqnarray}
where we have used the above topological formula again.
Let us note here that for high orders of perturbation theory (\ref{III6})
shows convergence only for $d$ very close to $d=4$, a problem known
for $\epsilon$ expansions.

{\em StepN:} Let $\bar{\cal G}\in[{\cal G}_0]$ be a given graph, contributing
to $\Gamma_{{\cal G}_0}$. We may assume (\ref{III6}) for all star subgraphs
${\cal G}\prec\bar{\cal G}$ with respect to their star subgraphs
$g_0\prec{\cal G}$. Let $R$ be any important subgraph of $\bar{\cal G}$.
As in {\em Step0}, the important connected component $\bar{g}_0$ of $R$
is an important star subgraph $\bar{g}_0\prec\bar{\cal G}$ 
and it is sufficient to show (\ref{III6}) for $\bar{g}_0$ and $\bar{\cal G}$.
We rescale the variables of $\bar{g}_0$ in
$\tilde{J}_{\bar{\cal G}}^{\rm ps *}$ by a factor $\rho^2$, decompose
$\tilde{J}_{\bar{\cal G}}^{\rm ps *}$, and reinsert the recursion for
$\tilde{J}_{\bar{g}_0}^{\rm ps *}$. Then, for the sum of inverse power
polynomials in $\tilde{J}_{\bar{\cal G}}^{\rm ps *}$, we find
\begin{eqnarray}\label{III7}
\rho^{dL_{\bar{g}_0}-2} \Delta_{\bar{\cal G}}^{\rm ps} &=&
\rho^{dL_{\bar{g}_0}-2} \left[ \Delta_{\bar{\cal G}}^{-\frac{d}{2}}
- \Delta_{\bar{\cal G}/\bar{g}_0}^{-\frac{d}{2}}
  \Delta_{\bar{g}_0}^{-\frac{d}{2}}\right]
\\ \nonumber
&& -\sum_{g\prec \bar{g}_0} \rho^{dL_{\bar{g}_0}-2}
\left[\Delta_{\bar{\cal G}/g}^{-\frac{d}{2}}
-\Delta_{\bar{\cal G}/\bar{g}_0}^{-\frac{d}{2}}
 \Delta_{\bar{g}_0/g}^{-\frac{d}{2}}\right] \Delta_g^{\rm ps}
\\ \nonumber
&& - \sum_{{g\prec\bar{\cal G}}\atop{g \not\subseteq \bar{g}_0}}
\rho^{dL_{\bar{g}_0/(\bar{g}_0\cap g)}}
\rho^{dL_{\bar{g}_0\cap g}-2}
\Delta_{\bar{\cal G}/g}^{-\frac{d}{2}}\Delta_g^{\rm ps}
\quad.
\end{eqnarray}
For important subgraphs $g,\bar{g}\prec{\bar{\cal G}}$ also the intersection
is an important star subgraph $g\cap\bar{g}\prec{\bar{\cal G}}$.
Therefore, scaling all variables of $\bar g_0$ by $\rho^2$,
the last term is regular at $\rho=0$ because of our inductive assumption.
For the other two terms we argue as in {\em Step1}, to find that
the right hand side of (\ref{III7}) is regular at $\rho=0$.
To include the exponential prefactor and possible subtractions, one may
replace all $\Delta_g^{-\frac{d}{2}}$ by $\tilde{J}_g^{\rm *}$ and
$\Delta_g^{\rm ps}$ by $\tilde{J}_g^{\rm ps *}$, to find the same result.
We recall the definition of the degree of divergence
$\omega_{\bar{g}_0}(\tilde{J}_{\bar{\cal G}}^{\rm ps *})$
as the number $\omega_{\bar{g}_0}$ for which
with appropriate ordering of variables
\begin{equation}\label{III8}
\rho^{\omega_{\bar{g}_0}+2\ell_{\bar{g}_0}}
\tilde{J}_{\bar{\cal G}}^{\rm ps *}
(\rho^2s_1,\ldots,\rho^2s_{\ell_{\bar{g}_0}},
s_{\ell_{\bar{g}_0}+1},\ldots,s_{\ell_{\bar{\cal G}}}) =
c_0 + {\cal O}(\rho) \quad\mbox{with}\quad c_0\neq 0
\quad.
\end{equation}
Using this,
we have shown (\ref{III6}) for $\bar{g}_0$ and $\bar{\cal G}$ and made
the inductive step forward.
This concludes our proof for the existence of the short chain limit
in $d<4$ for all partial sums $\tilde I_{\cal G}^{\rm ps}$, contributing to
the factor $\tilde{\cal C}_{{\cal G},{\cal A}}$ of the factorization
of the network vertex function $\tilde \Gamma_{{\cal G}_0}$.
We will now have to proceed, to see that the result persists in
analytical continuation and renormalization near 
the upper critical dimension $d=4$ of the theory.
\subsection{Renormalization}
Up to this point, we have shown that the integrals for the partial sums
of amplitudes $\tilde I_{\cal G}^{\rm ps}$ exist in the short chain limit
below $d=4$.
Using the dimensional scheme described in appendix \ref{A3}, we may
then construct the analytical continuation of $\tilde I_{\cal G}^{\rm ps *}$
to a strip $4<d<d^+$ above $d=4$. For the propagator subgraphs we
already assumed such a continuation from below $d=2$.
We can then renormalize the amplitudes $\tilde I_{\cal G}^{\rm ps *}$
near $d=4$, by extracting the poles at $d=4$ for all subgraphs of
$\cal G$. To keep track of multiple subtractions, this is performed
in a Bogoliubov recursive way \cite{Bogoliubov59}
 or equivalently by a Zimmermann
forest formula \cite{Zimmermann70}. In the scheme of Berg\`ere and David
\cite{Bergere81}, this extraction of poles subtracts from the integrand
of $\tilde I_{\cal G}^{\rm ps *}$ products of distributions in Schwinger
parameter space. In appendix \ref{A4},\ref{A5} we show that these exist also
for the Laplace transformed integrals
$I_{\cal G}^{\rm ps *}$.
Apart from the problem that some mass parameters of the integrand of
$\tilde I_{\cal G}^{\rm ps *}$ vanish, it is nothing but a sum of products
of Feynman integrals. Thus, having cleared the problem of vanishing
mass, any usual renormalization procedure applies, and will renormalize
the amplitudes $\tilde I_{\cal G}^{\rm ps *}$ as well as their sum
$\tilde{\cal C}_{{\cal G}_0,{\cal A}}^*$.
We want to find here the structure of the corresponding renormalization
factor, to determine the
scaling properties of $\tilde{\cal C}_{{\cal G}_0,{\cal A}}^*$
as function of the `short' chains.
Any network vertex function $\Gamma_{{\cal G}_0}$ and star vertex
function $\Gamma_{{\cal G}_0/{\cal A}}$ are renormalized
multiplicatively by $Z$ factors \cite{Schaefer92}.
Before taking the short chain limit
$\tilde{\cal C}_{{\cal G}_0,{\cal A}}$
will be renormalized by their ratio.
Indicating renormalized quantities by an index `R',
\begin{eqnarray}\label{III9}
\tilde\Gamma_{{\cal G}_0,{\rm R}} &=&
Z^{-1}_{{\cal G}_0} \tilde\Gamma_{{\cal G}_0}\quad,
\nonumber\\
(\tilde{\cal C}_{{\cal G}_0,{\cal A}})_{\rm R} &=&
Z^{-1}_{{\cal G}_0,{\cal A}} \tilde{\cal C}_{{\cal G}_0,{\cal A}} ,
\quad Z^{-1}_{{\cal G}_0,{\cal A}}=
Z^{-1}_{{\cal G}_0}/Z^{-1}_{{\cal G}_0/{\cal A}}.
\end{eqnarray}
If the pole structure of $\tilde{\cal C}_{{\cal G}_0,{\cal A}}$
persists in the short chain limit, then $Z^{-1}_{{\cal G}_0,{\cal A}}$
will also renormalize $\tilde{\cal C}_{{\cal G}_0,{\cal A}}^*$.
Looking at the partial sums of amplitudes, it seems that the pole
structure might be changed by cancellation, but in fact this cancellation
applies as well to the renormalization subtractions such that,
with $R'$ denoting the operation of renormalization of integrals
\begin{equation}\label{III10}
 R'(\tilde I_{\cal G}^{\rm ps}) = (R'\tilde I_{\cal G})^{\rm ps},
\end{equation}
the renormalized partial sum is equal to the partial sum of renormalized
amplitudes.
Following the above argument, the short
chain limit of the left hand side exists.
This limit of vanishing mass parameters for the `long' chains does
not imply any change of degree of divergence for any subgraph of
$\cal G$ with respect to $\tilde J_{\cal G}^{\rm ps}$, such that
all renormalization subtractions are kept, and no additional
subtractions occur in this limit.
The pole structure persists in the limit and the $Z$ factor remains
unchanged. Therefore, we can extract the scaling properties of
${\cal C}_{{\cal G}_0,{\cal A}}^*$ as function of the short chains 
from the ratio of the network and the star $Z$ factors.
For the exponent of the corresponding power law a scaling relation
results:
\begin{equation}
 {\cal C}^*_{{\cal G}_0,{\cal A}}(S) \sim S^{\nu\Theta_{{\cal G}_0,{\cal A}}}
,\quad \nu\Theta_{{\cal G}_0,{\cal A}} =
\gamma_{{\cal G}_0} - \gamma_{{\cal G}_0/{\cal A}}\quad,
\end{equation}
where the $\gamma$ exponents are the configuration number exponents
of the network ${\cal G}_0$ and of the star ${\cal G}_0/{\cal A}$
as defined in (\ref{I4}).

The proof of this relation, together with the proof that the
limit ${\cal C}_{{\cal G}_0,{\cal A}}^*$ is well defined,
is the central result of this paper.
We will discuss its consequences in the following chapter.
%
\section{Consequences and Conclusion}\label{IV}
\label{Kap15}
\subsection{OPE on a fractal: The SCE for polymer networks and stars}
\label{IVa}
For any network ${\cal G}_0$ with a subnetwork $\cal A$ of short chains,
we have proven in the preceding chapters that the vertex part of the
partition sum, and thus the partition sum itself, will factorize
in the limit of short chains $S_1\ll S_2$ - see note after (\ref{III2}):
\begin{equation}
 {\cal Z}_{{\cal G}_0}(S_1,S_2) \sim
 {\cal C}^*_{{\cal G}_0,{\cal A}}(S_1)
 {\cal Z}_{{\cal G}_0/{\cal A}}(S_2)
\quad.
\end{equation}
Here $S_1$ and $S_2$ define the chain length of the short and long
chains in the continuous model.
We have shown that this limit exists for the factor
${\cal C}^*_{{\cal G}_0,{\cal A}}$ and from its renormalization properties
we derived the scaling law and relation of exponents
\begin{equation}
{\cal C}^*_{{\cal G}_0,{\cal A}}(S) \sim
S^{\nu\Theta} \quad,\quad
\nu\Theta = \gamma_{{\cal G}_0} - \gamma_{{\cal G}_0 / {\cal A}} .
\end{equation}
Let us first discuss the consequences for two simple examples: A polymer star
of $F$ chains, one of which is short, and as a second example, a network
of two stars connected by one short chain, see fig.\ref{figIb}.
Let ${\cal G}_0$ be a star with $F$ arms, and the subnetwork ${\cal A}=\{ F \}$
be the set containing the short arm $F$.
Then the partition sums are the partition sums of the $F$ star
${\cal Z}_{{\cal G}_0}={\cal Z}_F$ and the $F-1$ star
${\cal Z}_{{\cal G}_0/{\cal A}}={\cal Z}_{F-1}$,
and with $S_1$ and $S_2$ the lengths of the short and long chains,
\begin{equation}
{\cal Z}_{{\cal G}_0}(S_1,S_2) \sim
S_1^{\gamma_F - \gamma_{F-1}}
{\cal Z}_{{\cal G}_0/{\cal A}}(S_2)\quad,
\end{equation}
where $\gamma_f$ is the configuration number exponent of the $f$ star
defined in chapter \ref{Ia}.

Let us now discuss a network ${\cal G}_0$ with a $f_1$ vertex and a $f_2$
vertex connected via one short chain, the other $f_1+f_2-2$ long
chains dangling from the open legs of the two vertices.
For this network the scaling relation in (\ref{I4}) gives the configuration
number exponent in terms of star exponents by:
\begin{equation}
\gamma_{{\cal G}_0} -1 = (\gamma_{f_1}-1) + (\gamma_{f_2}-1) - (\gamma-1)
\quad.
\end{equation}
After contracting the short chain of $\cal A$ to a vertex, the network
${\cal G}_0/{\cal A}$ is a $f_1+f_2-2$ star and so we find with the
same notation as above
\begin{equation}
{\cal Z}_{{\cal G}_0}(S_1,S_2) \sim S_1^{\nu\Theta}
{\cal Z}_{{\cal G}_0/{\cal A}}(S_2)
\quad,\quad
\nu\Theta = \gamma_{f_1}+\gamma_{f_2} - \gamma_{f_1+f_2-2} -\gamma
\quad.
\end{equation}
The $\gamma$ here is reminiscent of the fact, that in the original
network the two vertices were connected by one polymer chain
and in this way the law differs from the kind of relation resulting
for the contact exponents. Indeed we are expanding not just for short
distance between the two vertices but for short `distance' along
the fractal perimeter of the chain connecting the two vertices.
The fact that this polymer chain is a non trivial fractal,
introduces the additional $\gamma$. 

In the corresponding field theory the vertices are described by
composite operators $\phi_{f_1}$, $\phi_{f_2}$. So we can think of
our expansion as an OPE on a fractal, the polymer chain connecting
the two vertices. In fact, when transforming to field theory we are
also giving up the notion of chain lengths $S$ in favor of chemical
potentials $\mu$, which are mass parameters in the field theoretical
formulation. The SCE in this way corresponds to a large mass expansion
in field theory.
\subsection{Osmotic pressure for short and long chains}\label{IVb}
Solutions of long and short polymer chains chains 
\cite{Krueger89,Witten82,Krueger91} are also accessible experimentally.\\
The behavior of the osmotic pressure $\Pi$ for low concentrations may be
described by a virial expansion as in (\ref{I6}).
The mixed second virial coefficient $A^{(2)}_{12}$ describes the
interaction between the short and the long chains.
For dilute solutions we can restrict ourselves to the interaction of one
short chain with one long chain.
We find the following representation \cite{Krueger89} in the continuous
chain model
\begin{equation}
\label{2.1}
A^{(2)}_{12}(S_1,S_2) = 
\frac{{\cal Z}_{1,2}(S_1,S_2)}{{\cal Z}(S_1){\cal Z}(S_2)}\: .
\end{equation}
Here, ${\cal Z}(S_1)$, ${\cal Z}(S_2)$ are the partition sums of the single
chains, and ${\cal Z}_{1,2}(S_1,S_2)$ is the partition sum of the system of 
two chains
with chain lengths $S_1$ and $S_2$.
Interpreting one interaction point of the two chains as a 4-star vertex,
we can relate this partition sum to the 4-star partition sum.
To this end, we describe the system of the short and the long chain in solution
as a ternary polymer system, with couplings $u^{(1,1)}, u^{(2,2)}$
on the chains 1 and 2, and the coupling $u^{(1,2)}= u^{(2,1)}$
between the two chains 1 and 2.
The partition sum is then written as
${\cal Z}_{1,2}(S_1,S_2;u^{(a,b)})$
for symmetric couplings
${u^{(1,1)}= u^{(1,2)}=u^{(2,2)}= u}$.
The partition sum ${\cal Z}_{1,2}(S_1,S_2)$
may now be derived from the partition sum
${\cal Z}_4$ of a 4 arm polymer star \cite{Schaefer91}.
This can be easily shown, looking at the graphical expansion of
${\cal Z}_{1,2}$. In every connected graph each inter-chain interaction is
interpreted in turn as a 4-star vertex.
Formally this correspondence is given by the partial derivative of
${\cal Z}_{1,2}$ in $u^{(1,2)}$ :
\begin{eqnarray}\label{IV2-2}
\frac{\partial}{\partial u^{(1,2)}}{\cal Z}_{1,2}(S_1,S_2;u^{(a,b)})
\big|_{u^{(a,b)}=u} =
&&\int_0^{S_1}  ds_1 \int_0^{S_2}  ds_2
{\cal Z}_4 (s_1,S_1-s_1,s_2,S_2-s_2;u)  \:.
\end{eqnarray}
From the short chain expansion we know that for
$\frac{S_1}{S_2} \to 0$:
\begin{equation}
\label{5.3}
\int_0^{S_2}{\rm d}s_2 \:
{\cal Z}_4 (s_1,S_1-s_1,s_2,S_2-s_2;u)
\begin{array}[t]{c}
{\scriptstyle {\longrightarrow} }
\\[-1.0 ex] {\scriptstyle {\frac{S_1}{S_2}\to 0} }
\end{array}
{\cal C}^*(s_1,S_1-s_1;u)\: S_2 \: {\cal Z}(S_2) \quad,
\end{equation}
neglecting end effects of the integration along $S_2$.
Reintegrating (\ref{IV2-2}) we receive the form suggested in
\cite{Witten82},
\begin{equation}
\label{5.4}
A^{(2)}_{12}(S_1,S_2) \begin{array}[t]{c}
{\scriptstyle {\longrightarrow} }
\\[-1.0 ex] {\scriptstyle {\frac{S_1}{S_2}\to 0}}
\end{array}
\frac{S_2}{S_1}\frac{S_1 \int_0^{u} du'
\int_0^{S_1} ds {\cal C}^*(s,S_1-s;u')}{{\cal Z}(S_1)}  \;,
\end{equation}
with a linear dependence on $S_2$ as expected also from the blob
picture (fig. \ref{Ib}).
\subsection{Correlations of a multifractal}\label{IVc}
The flux of harmonic diffusion onto an absorbing fractal
defines a multifractal measure \cite{Cates87}.
The solution of the diffusion equation may be described in terms
of random walks using the path integral formalism.
If the absorber itself is a polymer, then the random walks, describing
diffusing particles, and the selfrepelling walk, describing the absorber,
may be incorporated in the same formalism. The flux to a segment
of the polymer is described by random walks with one end at that segment
and avoiding the polymer for the rest of the walk.
This mapping is analyzed in close detail in a forthcoming paper
\cite{FerHol96}.
Here we will use the fact that the $m$ th moment of the flux is given
by the number of configurations of $m$ random walks, ending at the
same segment on the polymer. Taking the average over all polymer
configurations as well, this is the partition sum of a polymer star,
consisting of $m+2$ arms of which $m$ are random walks, avoiding
the other two self and mutually avoiding walks.

With appropriate normalization, the flux onto the polymer defines
a multifractal measure $\mu$ and its moments $\langle \mu^n \rangle$.
Correlations of two moments $\langle \mu^m \mu^n \rangle$ are then
to be calculated from the partition sum of a network, consisting of
two vertices of the kind described above,  with $f_1=m+2$ and $f_2=n+2$
legs connected via the polymer 'backbone' of the two stars, just
as in our example above.
For short distance along the fractal carrier of the measure, that
is for a short chain of the polymer connecting the two vertices, we
may conclude from our proof, that this partition sum will factorize
into the partition sum for the $(m+n)$th moment and a factor depending
on the short chain length alone.
In fact, we assume here that our proof needs no essential change
to include also the random walks avoiding the polymer.
Using the scaling relations for polymer network exponents,
in terms of polymer star exponents, we conclude for the correlations
of the multifractal measure defined from the flux of harmonic diffusion,
\begin{equation} \label{eq:5.2.1}
\langle \mu^m(x) \mu^n(x+r) \rangle \sim {r}^{\Theta}
\langle \mu^{m+n}(x) \rangle
,\quad
\Theta = d_m + d_n - d_{m+n} \quad,
\end{equation}

the relation found by Cates and Deutsch \cite{Cates87a}.
The  problem of the multifractal flux of harmonic diffusion to an
absorbing polymer is mapped in detail to polymer field theory
in \cite{FerHol96}, discussing also the appropriate operators which
will produce the multifractal spectrum of exponents.
For these operators in fact eq. (\ref{eq:5.2.1}) is an operator
product expansion on the fractal carrier of the measure.
\subsection{Conclusions}\label{IVd}
The OPE of Wilson \cite{Wilson69} and Zimmermann \cite{Zimmermann70}
which expands the product of operators at
$x$ and $x+r$ in a series of composite operators at $x$ for $r\to 0$,
leads to the series of des~Cloizeaux' contact exponents in polymer theory.
We have established here
an analogous short chain expansion for networks consisting
of long chains $S_2$ and short chains $S_1$.
The existence of the first term of the expansion we have shown here to all
orders of renormalized perturbation theory.
This gives a basis for the derivation of scaling laws for systems of long and
short chains.
We could here treat the short chain expansion for quite a general class of
networks ${\cal G}_0$, but in pulling together the short chain
subnetwork $\cal A$, a star ${\cal G}_0/{\cal A}$ of long chains
has to result.
The case of loops in ${\cal G}_0/{\cal A}$ poses the problem that in the
perturbative treatment which we have adopted here, interactions between short 
chains and the long loop cannot be associated with the vertex, as in the 
star case.
By arguments derived from the short chain expansion, one might argue that
these contributions vanish in the short chain limit.
In summary, we have established an operator product expansion
(OPE) which, in terms
of polymer theory, is an expansion for short chain lengths between the
network vertices, described by the composite operators.
We have used this expansion as a tool to analyse the interaction of long
and short polymers in solution, as measured via the virial coefficients
of the osmotic pressure.
Mapping the two interacting chains to a $4$ star polymer, we found
a linear dependence
of the second virial coefficient on the long chain length.
This confirms a claim footing on a blob argument.\\
On the other hand, our method gives new insight into the problem of
correlations of a multifractal measure. For the measure defined by the
flux of harmonic diffusion onto an absorbing polymer we could describe
the correlations of two moments $\langle \mu^n\mu^m\rangle$ by a corresponding
polymer and random walk network. The SCE then gives the answer for
short distance correlations on the fractal carrier of the measure
again confirming the blob result.
\acknowledgements
I would like to thank L. Sch\"{a}fer for introducing me to the 
subject, for numerous discussions and valuable advice. 
This work has been supported by the Deutsche Forschungsgemeinschaft,
SFB 237, `Unordnung und gro\ss{}e Fluktuationen'.
\begin{appendix}
\section{Graph Contributions and their Renormalization}\label{A}
\subsection{Graphs and their amplitudes}\label{A1}
\begin{figure}
\epsfbox{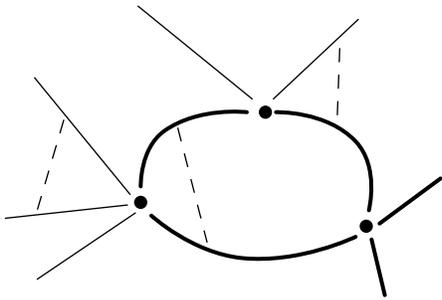}
\caption{\label{figAa}
A graph contribution $\cal G$ to
the vertex part $\Gamma_{{\cal G}_0}$ of the partition function
of the network ${\cal G}_0$ shown in fig. 2.}
\end{figure}

The contributions of perturbation theory to the partition function
${\cal Z}_{{\cal G}_0}$ or its vertex part $\Gamma_{{\cal G}_0}$
for a given network ${\cal G}_0$ are constructed from the graph
representing ${\cal G}_0$. The interactions, denoted by dotted
lines in Fig. \ref{figAa} are four leg vertices which in this `faithful'
representation (see e.g. \cite{Amit84}\S6.8) must preserve the topology
of the original network. To define the amplitude of the contribution
of the graph $\cal G$, we cut away any dangling ends, number all vertices
(original ${\cal G}_0$ vertices and interaction vertices) by
$j=1,\ldots,v_{\cal G}$ and number all lines connecting these vertices by
$a=1,\ldots,\ell_{\cal G}$; the set of lines we often denote by $\cal G$.
To every line $a$ a `mass' -parameter $m_a$ and a $d$ - dimensional
integration variable, `momentum', $k_a\in I\!\!R^d$ is associated.
For every ${\cal G}_0$ - vertex $j$ an external momentum $p_j$ may
occur.
For every chain $f\in{\cal G}_0$ of the original network, we form the
set $A_f\subset{\cal G}$ of lines which constitute this chain as a
subset of $\cal G$ and define $m_a^2=\mu_f$ for all $a\in A_f$.
Then, the amplitude $\tilde I_{\cal G}$ of the graph $\cal G$ is
defined as
\begin{equation}\label{eqA1}
\delta^d(\sum_{j=1}^{v_{\cal G}} p_j)\tilde I_{\cal G}(m_a^2,p_j)=
\int\prod_{a\in{\cal G}}\frac{{\rm d}^d k_a}{(2\pi)^{d/2}}
\frac{\vartheta_\Lambda(k_a)}{k_a^2+m_a^2}
\prod_{j=1}^{v_{\cal G}}(2\pi)^{d/2}\delta^d(
\sum_{j=1}^{v_{\cal G}}\sum_{a\in{\cal G}}\epsilon_{aj}k_a -p_j)
\end{equation}
Here $\vartheta_\Lambda$ is a regularizing function e.g.
$\vartheta_{\Lambda}(k)=\Theta(\Lambda^2-k^2)$. The incidence matrix
$\epsilon_{aj}$ indicates if a line $a$ begins or ends in the vertex $j$:
\begin{equation}
\epsilon_{aj}=\left\{ \begin{array}{rl}
 1 & a \mbox{ begins in } j \\
-1 & a \mbox{ ends in } j\\
 0 & a \mbox{ begins and ends in } j\\
 0 & \mbox{ else }.
\end{array}
\right.
\end{equation}
The direction of all lines is arbitrary, but fixed.
For vanishing external momenta $p_j=0$, we write the amplitude in
Schwinger parametrization \cite{Nakanishi71},
\begin{equation}
\tilde I_{\cal G} = \int_0^\infty \prod_{a\in{\cal G}}{\rm d}s_a
 \tilde J_{\cal G}(s),
\quad \tilde J_{\cal G}(s)=\vartheta(s)\exp(-\sum_{a\in{\cal G}}m_a^2s_a)
\Delta_{\cal G}(s)^{-d/2}
\quad,
\end{equation}
with a regularizing function $\vartheta(s)$ - see below - and the Symanzik
polynomial
\begin{equation}
\Delta_{\cal G}(s)=\prod_{a\in{\cal G}}s_a
\det\{\sum_{a\in{\cal G}}\frac{\epsilon_{ai}\epsilon_{aj}}{s_a}\}_{ij}
= \sum_T \prod_{a\in{\cal G}\setminus T}s_a .
\end{equation}
In the last representation the sum is over all trees $T$ on $\cal G$,
that is all subsets $T\subset {\cal G}$ of $v_{\cal G}-1$ lines, which
connect all vertices when $\cal G$ is connected.
There are many other characterizations
e.g. $T$ are maximal with no loops etc. \cite{Nakanishi71} .
One shows that $\Delta_{\cal G}$ is a homogeneous polynomial with degree
equal to the number of loops of $\cal G$:
$L_{\cal G} = \ell_{\cal G} - v_{\cal G} + r$, where $r$ is the number of
connected components of $\cal G$.
If $g$ is a subgraph of $\cal G$, then the sum over trees may be decomposed
into a sum over trees $T$ on $\cal G$ which are also trees on $g$ and
${\cal G}/g$, and all other trees $T'$. We receive
\begin{equation}
\Delta_{\cal G}=\Delta_{{\cal G}/g}\Delta_g + {\cal P}_{{\cal G},g},
\quad
{\cal P}_{{\cal G},g} = \sum_{T'} \prod_{a\in{\cal G}\setminus T'}s_a
\quad.
\end{equation}
Trees $T'$ on $\cal G$ which are not trees on $g$ will have less lines
on $g$ than a tree on $g$, such that ${\cal P}_{{\cal G},g}$ will be
of higher degree in the variables $s_a$, $a\in g$ than the polynomial
$\Delta_g$.
\subsection{Topological relations}\label{A2}
For a graph $\cal G$ constructed from the plain network graph ${\cal G}_0$
let $n_f$ be the number of vertices of $\cal G$ with $f$ legs. We consider
amputated graphs with $n_1=0$. The total number of vertices is $v_{\cal G}$.
For a connected graph the number of lines is $\ell_{\cal G}$:
$2\ell_{\cal G}=\sum_{f>1}n_f * f - q_{\cal G}$ with $q_{\cal G}$
the number of (amputated) external legs of $\cal G$.
The number of loops is $L_{\cal G}= \ell_{\cal G}-v_{\cal G}+1$
and the degree of divergence with respect to $\Delta_{\cal G}^{-d/2}$
is $\omega_{\cal G}= d L_{\cal G} - 2\ell_{\cal G}$.
If $g$ is a connected subgraph of $\cal G$ that contains all ${\cal G}_0$ -
vertices, then $2(\ell_{\cal G} - \ell_g) = 4(v_{\cal G}-v_g) -
(q_{\cal G}-q_g)$, as the two graphs only differ in the 4-functional
interaction vertices. The degrees of divergence with respect to
$\Delta_{\cal G}^{-d/2}$ differ by
\begin{eqnarray}
\omega_{\cal G}-\omega_g &=& d(L_{\cal G}-L_g) - 2(\ell_{\cal G}-\ell_g)
\nonumber\\
&=& d\left[(\ell_{\cal G}-\ell_g)-(v_{\cal G}-v_g)\right]
    - 4(v_{\cal G}-v_g) + (q_{\cal G}-q_g)
\nonumber\\
&=& (1-\frac{d}{2})(q_{\cal G}-q_g) - (4-d)(v_{\cal G}-v_g)
\quad.
\end{eqnarray}
\subsection{Regularization and renormalization}\label{A3}
We summarize here some points of the
dimensional scheme in the formulation of Berg\`ere,
David and Lam \cite{Bergere76,Bergere79,Bergere81} as far as they are
relevant for our argument.
We use multidimensional regularization which introduces a dimensional
parameter $h_g$ for every subgraph $g$ of $\cal G$:
\begin{equation}
\vartheta(s) = \prod_{g\subseteq{\cal G}} \Delta_g^{-h_g/2}(\{s_a\}_{a\in g})
\end{equation}
For graphs $\cal G$ constructed by the above procedure $I_{\cal G}$ will
converge for sufficiently small values of $d$ and $h_g$; i.e., for
$h_g=0$ below $d=2$.
$I_{\cal G}$ may be analytically continued in $d$ to a meromorphic function
in $d$ and $h_g$ with possible poles at $d=2$ and $d=4$. 
Let us define this procedure in some more detail. 
Assume the integrand
$\tilde J_{\cal G}(s_1,\ldots,s_{\ell_{\cal G}})$
to be any integrand occurring in chapter \ref{III}, including
$\tilde J_{\cal G}^{\rm ps *}$. Let $g=\{1,\ldots,\ell_g\}$ be a `subgraph'
of $\cal G$, defining a subset of integration variables.
We then define a generalized Taylor operator $\tau_g$ by
\begin{equation}
\tau_g \tilde J_{\cal G}(\{s_a\}) =
\sum_{k=0}^{[\omega_g]} \frac{1}{k!}
\left(\frac{\partial}{\partial\rho}\right)^k_{\rho=0}
\rho^{2\ell_g+\omega_g}
\tilde J_{\cal G}(\rho^2s_1,\ldots,\rho^2s_{\ell_g},
            s_{\ell_g+1},\ldots,s_{\ell_{\cal G}})
\end{equation}
where $\omega_g$ is the number for which the $k=0$ term is finite, and
$[\omega_g]\leq\omega_g$ is the nearest whole number below $\omega_g$.
When all $h_{g'}=0$ this defines the {\em degree of divergence}
$\omega_g(\tilde J_{\cal G})$ with respect to the integrand 
$\tilde J_{\cal G}$. If $\omega_g< 0$ then $\tau_g\tilde J_{\cal G}=0$,
 if $\omega_g\geq 0$
the subgraph $g$ is {\em divergent}. For integrands such as 
$\Delta_{\cal G}^{-d/2}$ the Taylor
operators $\tau_g$ will depend on dimension $d$.
The analytical continuation of the integral $\tilde I_{\cal G}$ of
$\tilde J_{\cal G}$ for a given dimension $d$ is constructed by replacing
\begin{equation}\label{eqA6}
\tilde J_{\cal G} \to R\tilde J_{\cal G} =
\prod_{g\subseteq{\cal G}}(1-\tau_g) \tilde J_{\cal G}
= \left[ 1 + \sum_{\cal F}\prod_{g\in{\cal F}}(-\tau_g)\right]
\tilde J_{\cal G}
\quad.
\end{equation}
Here the regularizing dimensions $h_g$ are chosen such as not to change the
degree of the Taylor operators. The sum in the second representation
of $R$ is over all `forests', sets $\cal F$ of non overlapping divergent
subgraphs: $g_1,g_2\in{\cal F}$, then either $g_1\cap g_2=\emptyset$
or $g_1\subset g_2$ or $g_2\subset g_1$.

To {\em renormalize} $I_{\cal G}$ near $d^*=4$, we can construct analytical
continuations in a strip $d^-<{\rm Re}(d)<d^*$ below $d^*$, with operators
$\tau_g=\tau_g^-$ in (\ref{eqA6}), and in a strip $d^*<{\rm Re}(d)<d^+$ above
$d^*$, using operators $\tau_g=\tau_g^+$.
Then we can extract the pole singularity at $d=d^*$ by Cauchy integration
around $h_g=0$ for every subgraph $g$. For appropriate bookkeeping of
multiple subtractions this has again to be performed in a Bogoliubov
recursive way or by a forest formula.
We perform these integrations here using distribution formulas. Define
\begin{equation}\label{eqA7}
\tau'_g \tilde J_{\cal G}= \tau^-_g \tilde J_{\cal G} +
(\tau^+_g - \tau^-_g) \tilde J_{\cal G} \Theta(x_g)\;.
\end{equation}
Then the dimensionally renormalized integrand is represented by
\cite{Bergere81}
\begin{equation}\label{eqA11}
\tilde J_{\cal G} \to R'\tilde J_{\cal G} =
\left[ \tilde J_{\cal G} + \sum_{\cal F}\prod_{g\in{\cal F}}(-\tau'_g)
\tilde J_{\cal G}\right]_{\Delta'}
\end{equation}
The sum is again over all forests $\cal F$ of non overlapping divergent
subgraphs. The index $\Delta'$ stands for the substitution
$h_g \to \partial/\partial x_g|_{x_g=0}$, this operation on the
$\Theta$ - functions in (\ref{eqA7}) represents by distributions
the Cauchy integrations in the variables $h_g$.
\subsection{Dimensional renormalization in polymer variables}\label{A4}
This problem has also been adressed in \cite{Duplantier86h}.\\
Constructing the analytical continuation of $I_{\cal G}$ the $\tau_g$
operators also act on the exponential factors in the masses $\mu_f$
of the integrand $\tilde J_{\cal G}$. One receives polynomials
${\cal P}_{\cal F}(\mu_f)$ in $\mu_f$. For every term of (\ref{eqA6})
they occur in the combination
\begin{equation}
\prod_{g\in{\cal F}}(-\tau_g)
\prod_f \exp[-\mu_f\sum_{a\in A_f}s_a]\Delta_{\cal G}^{-d/2}(s)
= {\cal P}_{\cal F}(\mu_f;s)
\prod_f \exp[-\mu_f\sum_{a\in A_f\setminus{\cal F}}s_a]
\end{equation}
Here $ A_f\setminus{\cal F}$ denotes the set of lines $a$ of the chain $A_f$
which are not contained in any $g\in{\cal F}$. Laplace transforming
$\mu_f \to S_f$ we receive for $R J_{\cal G}(s,S)$ products of $\Theta^{(j)}$
functions of the form $\Theta^{(j)}(S_f-\sum_{a \in A_f\setminus{\cal F}}s_a)$
for every forest $\cal F$. The same product of distributions occur when
formally applying the $\tau_g$ operators on the $\Theta$ - functions in
$J_{\cal G}$. Thus the $R$ - operation or analytical continuation
obviously commutes with the Laplace transform.
The existence of these products of distributions together
with the product of distributions generated by the $R'$ renormalization
operation is shown in the next section.
\subsection{Distributions}\label{A5}
We investigate products of $\Theta$ functions in the integrals for the
Laplace transformed analytically continued and renormalized amplitudes.
The integrals $R\tilde J_{\cal G}$ of the analytically continued amplitudes
contain for every forest ${\cal F}$ a product of $\Theta$ functions of the form
\[
     \prod_{a=1}^{\ell} \Theta(s_a) \prod_{f=1}^{F}
    \Theta^{(j_f)}(S_f - \sum_{a\in A_f\setminus{{\cal F}}}
s_a ) ,
\]
In the integrals of the renormalized amplitudes we have
\[
  \prod_{a=1}^{\ell} \Theta(s_a) \prod_{f=1}^{F}
    \Theta(S_f - \sum_{a\in A_f\setminus{{\cal F}}} s_a)
   \prod_{g\in {\cal F}} \Theta^{(j_g)}(-\frac{1}{2}\ln Q_{g,{\cal F}}(s)) .
\]
The $ \Theta^{(j_g)}(-\frac{1}{2}\ln Q_{g,{\cal F}}(s))$ distributions
result from the operation in (\ref{eqA11}), see \cite{Bergere81}.
The $Q_{g,{\cal F}}$ are products of polynomials
\begin{equation}
     Q_{g,{\cal F}} = 
     \prod_{{g'\in{\cal F}}\atop{g'\subset g}}
         \Delta_{g'/{\cal F}}   \;,
\end{equation}
where $g'/{\cal F}$ denotes the graph in which all $g'' \in {\cal F}$,
contained in $g'$, $g'' \subset g'$ have been contracted to vertices.

These products of distributions exist if the hypersurfaces defined by their
arguments have no tangent intersection.
This means the vectors
\begin{equation}
  \vec{n}_{g,{\cal F}} =
 \frac{\partial}{\partial\vec{s}}   \ln Q_{g,{\cal F}}(s) ,\quad
  \vec{n}_{A_f,{\cal F}} =
  \frac{\partial}{\partial\vec{s}}
         (\sum_{a\in A_f\setminus{{\cal F}}} s_a) , \quad
  \vec{n}_{a} =
  \frac{\partial}{\partial\vec{s}} (s_a)
 \end{equation}
must be linearly independent on arbitrary intersections of these surfaces.
This has been shown for the vectors $\vec{n}_{g,{\cal F}}$,$\vec{n}_{a} $
in \cite{Bergere81}.
For $\vec{n}_{A_f,{\cal F}}$ we have:
\begin{itemize}
\item[(1)]
As the $A_f$ are disjoint by definition, the $\vec{n}_{A_f,{\cal F}}$
are orthogonal.
\item[(2)]
The polynomials $Q_{g,{\cal F}}$ contain only those $s_a$
variables which belong to one $g'\in {\cal F}$.
\\  Only those lines $a$ contribute to
 $\sum_{a\in A_f\setminus{{\cal F}}}$
which belong to no $g'\in {\cal F}$.
\\  Therefore the vectors $\vec{n}_{g,{\cal F}} $ and
$\vec{n}_{A_f,{\cal F}}$ are orthogonal.
\item[(3)] If $\vec{n}_{A_f,{\cal F}}$ and $\vec{n}_a$ are linearly dependent
there is a subgraph $g_0$ with
\[
 \vec{n}_{A_f,{\cal F}} = \sum_{a\in g_0} \lambda_a \vec{n}_a \quad ,
 \quad \lambda_a \neq 0 \quad \mbox{for all} \quad a\in g_0  \;,
\]
and $ A_f\setminus{{\cal F}}$ must be contained in $g_0$.
Then the intersection of the hyperplanes described by $s_a=0$, $a\in g_0$,
with the hyperplane described by
$\sum_{a\in A_f\setminus{{\cal F}}} s_a=S_f$,
is empty for $S_f\neq 0$.
\[
      \bigcap_{a\in g_0} \{s:s_a=0\}\cap
                         \{ s :\sum_{a'\in A_f\setminus{{\cal F}}}
s_{a'} =S_f  \}
= \emptyset   \;.
\]
Therefore the vectors $\vec{n}_{A_f,{\cal F}}$ and $\vec{n}_a$
can be linearly dependent only where the hypersurfaces do not intersect.
\end{itemize}
\section{Amplitudes for Vanishing Masses}\label{B}
Consider an amplitude $\tilde I_{\cal G}^*$ in the short chain limit, in which
all mass parameters $\mu_f$ of `long' chains vanish:
\begin{equation}
\tilde{I}^*_{\cal G} = \int_0^\infty \prod_{a \in {\cal G}} d s_a
\prod_{f \in {\cal A}} {\rm e}^{- \mu_f
\sum_{a \in A_f} s_a} {Z}_{\cal G}(s)  \;,
\end{equation}
where the integrand $Z_{\cal G}(s)$  may be any of
the functions $\Delta_{\cal G}^{-d/2}$,
$\Delta_{\cal G}^{\rm ps }$, defined in chapter \ref{III}.
The integration area is divided in Hepp sectors, in each of which the
$s_a$ are ordered according to a permutation $\sigma \in {\wp}_l$:
$s_{\sigma_1}\le s_{\sigma_2}\le\ldots\le s_{\sigma_l}$
Here $l=l_{\cal G}$ is the number of lines of the graph ${\cal G}$.

Using the substitution $s_{\sigma_j}\to \prod_{i=j}^l \beta_i^2$
with $\beta_j^2=s_{\sigma_j}/s_{\sigma_{j+1}}$ and
$\beta_{\ell}^2=s_{\sigma_{\ell}}$ on each sector, we have
\begin{equation}
\tilde{I}^*_{\cal G} = \sum_{\sigma \in {\wp}_l}
\tilde{I}_{\cal G}^\sigma
\quad,
\end{equation}
\begin{equation}
{I}_{\cal G}^\sigma = \int_0^\infty \prod_{j \in {\cal G}}
d \Big( \prod_{i=j}^l \beta_i^2 \Big) \prod _{f \in {\cal A}}
{\rm e}^{- \mu_f \sum_{\sigma_{j} \in A_f}
\prod_{i=j}^l \beta_i^2}
{Z}_{\cal G} \Big(\Big\{ \prod_{i=\sigma^{-1}_a}^l \beta_i^2
\Big\} _a \Big)  \;.
\end{equation}
The substitution is chosen in such a way that integration in $\beta_i$,
$i<l$ is from 0 to 1 and  only in $\beta_l$ from 0 to $\infty$.

Let us now consider the sector $\sigma$ with
$s_1 \le \ldots \le s_l$. This is no restriction from generality.
\begin{equation}
\tilde{I}_{\cal G}^{\sigma} = 2^l \int_0^1 \prod_{a=1}^{l-1}
d \beta_a \, \beta_a^{-\omega_j-1}
\int_0^\infty d \beta_l \, \beta_l^{-\omega_l-1}
\prod_{f \in {\cal A}} {\rm e}^{-\mu_f  \sum_{a \in A_f}
\prod _{j=a}^l \beta_j^2} {Z}_{\cal G}^{\sigma}
(\{ \beta_a \}_{a<l} )  \;.
\end{equation}
Depending on the degree of divergence $\omega_j$ of the subgraph given by
$R_j^{(\sigma)} = \{ \sigma_1, \sigma_2, \ldots,
\sigma_j \} = \{ 1,2,\ldots, j\}$ with respect to ${Z}_{\cal G}$
corresponding powers of $\beta_j$ were
extracted from ${Z}_{\cal G}$,
\begin{equation}
{Z}_{\cal G} \Big( \Big\{ \prod _{i=\sigma_a^{-1}}^l
\beta_i^2 \Big\}_a \Big) = \prod _{j=1}^l
\beta_j^{-\omega_j -2j} {Z}_{\cal G}^\sigma
(\{ \beta_a \}_{a<l} )\quad,
\end{equation}
and were combined with the Jakobian of the substitution to powers
in the degree of divergence $\omega_j(\sigma)$.
The integral in $\beta_l$ may now be performed and we receive
\begin{equation}
\tilde{I}_{\cal G}^{\sigma} = 2^{l-1} \Gamma
\left( - \frac{\omega_l}{2} \right) \int_0^1 \prod_{j=1}^{l-1}
d\beta_j \, \beta_j^{-\omega_j-1} \left( \sum_{f \in {\cal A}}
\mu_f \sum_{a \in A_f} \prod_{i=a}^{l-1} \beta_i^2
\right)^{\frac{\omega_l}{2}} {Z}_{\cal G}^\sigma (\beta) \;.
\end{equation}
By construction the function ${Z}_{\cal G}^\sigma (\beta)$ is regular for
$\beta_j=0$. For the integrals to exist all subgraphs $R_j$ must be
convergent: $\omega_j <0$ for all $j \le l$. In addition we have to
investigate the polynomial $\sum_{f \in {\cal A}} \mu_f
\sum_{a \in A_f} \prod_{i=a}^{l-1} \beta_i^2$.
If $j_\sigma$ is the highest index which denotes a massive line
$\sigma_{j_\sigma} \in A_{f_\sigma}$, $f_\sigma
\in {\cal A}$ then the product
$\prod_{i=j_\sigma}^{l-1} \beta_i^2$ is contained in all terms of this
polynomial and we can write
\begin{eqnarray}
{I}_{\cal G}^\sigma = \int_0^1 \prod_{j=1}^{l-1} d\beta_j
\prod_{i=1}^{j_\sigma-1} \beta_i^{-\omega_i-1}
\prod_{i=j_\sigma}^{l-1} \beta_i^{\omega_l-\omega_i-1}
\left( \mu_{f_\sigma} + \sum_{f\in{\cal A}}\mu_f
\sum_{\sigma_j \in A_f}
\prod_{i=j}^{j_\sigma-1} \beta_i^2 \right)^{\frac{\omega_l}{2}}
{Z}_{\cal G}^\sigma (\beta) \;.
\end{eqnarray}
Obviously the integrals are finite if on all Hepp sectors all subsectors
$R_i$ are convergent, $\omega_i < 0$, and all subsectors $R_i$ containing
the massive subgraph $i \ge j_\sigma$ are more convergent than the total
graph $\cal G$, $\omega_i < \omega_l < 0$.

For $i \ge j_\sigma$ all massive chains $A_f$ are contained $R_i$ resp.
${\cal G} / R_i$ consists only of massless lines.
We define:
\begin{itemize}
\item[{\bf Def.}]
 A subgraph $g \subset {\cal G}$ is called important subgraph of ${\cal G}$,
if ${\cal G} /g$ contains massless lines only.
\end{itemize}
We have thus shown the following lemma:\cite{Bergere76a}
\begin{eqnarray}
\lefteqn{\tilde{I}^*_{\cal G} \mbox{ is finite if }} && \nonumber\\
&\mbox{(i)}&\mbox{All subgraphs $g \subset {\cal G}$ are convergent
$\omega_g < 0$ .}\nonumber\\
&\mbox{(ii)}&\mbox{All important subgraphs $g \subset {\cal G}$ are more
convergent
than $\cal G$, $\omega_g < \omega_{\cal G} < 0$.}
\label{A11.1}
\end{eqnarray}
\end{appendix}

\newpage
\center{\bf FIGURES}

\vspace{3ex}
\unitlength=1.00mm
\special{em:linewidth 0.4pt}
\linethickness{0.4pt}
\begin{figure}
\begin{eqnarray*}
\begin{array}[t]{c}
I_1: 
\begin{array}{c} 
\begin{picture}(16.17,8.28)
\emline{4.58}{5.81}{1}{5.58}{7.10}{2}
\emline{6.50}{7.96}{3}{8.33}{8.28}{4}
\emline{9.25}{8.28}{5}{10.58}{7.85}{6}
\emline{11.17}{6.99}{7}{12.08}{5.81}{8}
\emline{0.33}{4.73}{9}{16.17}{4.73}{10}
\put(15.58,3.33){\makebox(0,0)[rt]{$S$}}
\end{picture}
\end{array}
\\
I_2: 
\begin{array}{c}
\begin{picture}(16.17,7.74)
\emline{0.33}{4.73}{1}{16.17}{4.73}{2}
\emline{3.58}{5.27}{3}{4.58}{6.66}{4}
\emline{5.58}{7.42}{5}{6.42}{7.74}{6}
\emline{6.42}{7.74}{7}{7.17}{7.42}{8}
\emline{8.25}{6.66}{9}{9.17}{5.27}{10}
\emline{6.08}{3.98}{11}{7.42}{2.69}{12}
\emline{8.25}{2.15}{13}{9.25}{1.72}{14}
\emline{9.25}{1.72}{15}{10.67}{2.15}{16}
\emline{11.50}{2.69}{17}{12.75}{3.87}{18}
\end{picture}
\end{array}
\\
I_3: 
\begin{array}{c}
\begin{picture}(16.17,9.57)
\emline{0.33}{4.73}{1}{16.17}{4.73}{2}
\emline{5.75}{5.27}{3}{6.75}{6.66}{4}
\emline{7.75}{7.42}{5}{8.59}{7.74}{6}
\emline{8.59}{7.74}{7}{9.34}{7.42}{8}
\emline{10.42}{6.66}{9}{11.34}{5.27}{10}
\emline{3.67}{5.81}{11}{4.83}{7.42}{12}
\emline{6.17}{8.60}{13}{7.58}{9.57}{14}
\emline{9.42}{9.57}{15}{11.08}{8.60}{16}
\emline{12.25}{7.53}{17}{13.33}{5.81}{18}
\end{picture}
\end{array}
\end{array}
   &\hspace{3em}&
\begin{array}[t]{c}
\dot{I}_1^{ab}:
\begin{array}{c}
\begin{picture}(23.50,7.96)
\put(11.58,5.05){\circle*{0.50}}
\emline{1.58}{5.05}{1}{10.00}{4.95}{2}
\emline{13.33}{4.95}{3}{23.50}{4.95}{4}
\emline{5.92}{5.59}{5}{7.08}{6.77}{6}
\emline{8.08}{7.31}{7}{9.92}{7.85}{8}
\emline{11.58}{7.96}{9}{13.92}{7.53}{10}
\emline{14.83}{6.99}{11}{16.00}{5.70}{12}
\put(5.92,3.33){\makebox(0,0)[ct]{$S_a$}}
\put(18.33,3.33){\makebox(0,0)[ct]{$S_b$}}
\end{picture}
\end{array}
\\
I_4^{abc}: 
\begin{array}{c}
\begin{picture}(16.58,19.57)
\put(1.00,9.89){\circle*{0.94}}
\emline{2.33}{9.89}{1}{15.08}{9.89}{2}
\emline{1.75}{10.86}{3}{11.75}{19.57}{4}
\emline{1.83}{9.14}{5}{11.67}{1.08}{6}
\emline{9.58}{9.57}{7}{9.42}{8.06}{8}
\emline{9.17}{7.42}{9}{8.83}{6.24}{10}
\emline{8.50}{5.70}{11}{8.08}{4.73}{12}
\emline{9.92}{16.88}{13}{10.67}{15.48}{14}
\emline{11.17}{14.52}{15}{11.50}{13.01}{16}
\emline{11.67}{12.04}{17}{11.67}{10.54}{18}
\put(13.00,19.57){\makebox(0,0)[lc]{$c$}}
\put(16.58,9.89){\makebox(0,0)[lc]{$a$}}
\put(13.42,1.08){\makebox(0,0)[lb]{$b$}}
\end{picture}
\end{array}
\end{array}
\end{eqnarray*}
\caption{\label{figIe}
Graphs of the one loop integrals $I_1$, $\dot{I}_1^{ab}$
and two loop integrals $I_2$, $I_3$, $I_4$ occurring in the perturbation
expansion for the partition sum of a polymer star.}
\end{figure}
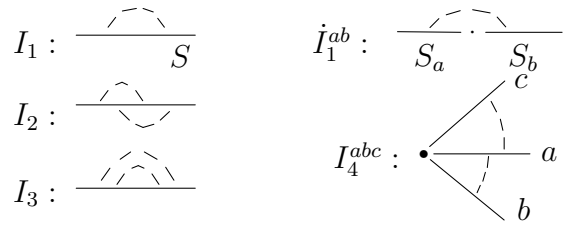
\begin{figure}
\[
I_{\cal G}^{ps} = 
\begin{array}{c}
\begin{picture}(16.58,19.57)
\put(1.00,9.89){\circle*{0.94}}
\emline{2.33}{9.89}{1}{15.08}{9.89}{2}
\emline{1.75}{10.86}{3}{11.75}{19.57}{4}
\emline{1.83}{9.14}{5}{11.67}{1.08}{6}
\emline{9.58}{9.57}{7}{9.42}{8.06}{8}
\emline{9.17}{7.42}{9}{8.83}{6.24}{10}
\emline{8.50}{5.70}{11}{8.08}{4.73}{12}
\emline{9.92}{16.88}{13}{10.67}{15.48}{14}
\emline{11.17}{14.52}{15}{11.50}{13.01}{16}
\emline{11.67}{12.04}{17}{11.67}{10.54}{18}
\put(13.00,19.57){\makebox(0,0)[lc]{$F$}}
\put(16.58,9.89){\makebox(0,0)[lc]{$a$}}
\put(13.42,1.08){\makebox(0,0)[lb]{$b$}}
\end{picture}
\end{array}
\; - \; (
\begin{array}{c}
\begin{picture}(16.67,10.65)
\put(1.08,0.97){\circle*{0.94}}
\emline{2.42}{0.97}{1}{15.17}{0.97}{2}
\emline{1.83}{1.94}{3}{11.83}{10.65}{4}
\emline{10.00}{7.96}{5}{10.75}{6.56}{6}
\emline{11.25}{5.59}{7}{11.58}{4.09}{8}
\emline{11.75}{3.12}{9}{11.75}{1.61}{10}
\put(13.08,10.65){\makebox(0,0)[lc]{$a$}}
\put(16.67,0.97){\makebox(0,0)[lc]{$b$}}
\end{picture}
\end{array}
\quad) \cdot (
\begin{array}{c}
\begin{picture}(16.67,10.65)
\put(1.08,0.97){\circle*{0.94}}
\emline{2.42}{0.97}{1}{15.17}{0.97}{2}
\emline{1.83}{1.94}{3}{11.83}{10.65}{4}
\emline{10.00}{7.96}{5}{10.75}{6.56}{6}
\emline{11.25}{5.59}{7}{11.58}{4.09}{8}
\emline{11.75}{3.12}{9}{11.75}{1.61}{10}
\put(13.08,10.65){\makebox(0,0)[lc]{$F$}}
\put(16.67,0.97){\makebox(0,0)[lc]{$a$}}
\end{picture}
\end{array}
\quad)
\]
\caption{\label{figIf}
Illustration of the partial sum $I_{\cal G}^{ps}$ for the three arm graph,
 in graphical notation.}
\end{figure}
\newpage
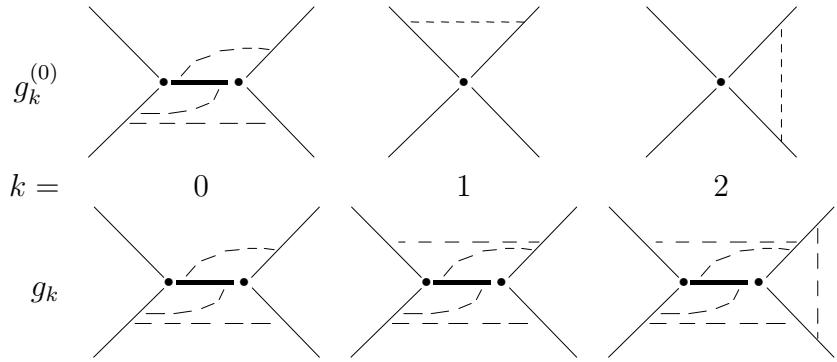
\begin{figure}
\[
\begin{array}{rccc}
g_k^{(0)}&
\begin{array}{c}
\begin{picture}(30.00,20.00)
\put(20.00,10.00){\circle*{0.93}}
\emline{9.33}{10.65}{1}{0.08}{20.00}{2}
\emline{9.33}{9.14}{3}{0.00}{0.00}{4}
\emline{21.08}{9.25}{5}{30.00}{0.00}{6}
\emline{21.00}{10.65}{7}{30.00}{20.00}{8}
\put(10.00,10.00){\circle*{0.93}}
\put(11.00,9.67){\rule{7.50\unitlength}{0.50\unitlength}}
\emline{7.00}{5.83}{9}{9.33}{5.83}{10}
\emline{9.33}{5.83}{11}{9.33}{5.83}{12}
\emline{10.67}{5.83}{13}{13.00}{6.17}{14}
\emline{14.33}{6.50}{15}{16.00}{7.17}{16}
\emline{16.67}{7.67}{17}{17.33}{9.00}{18}
\emline{12.33}{10.83}{19}{13.50}{12.17}{20}
\emline{14.33}{12.83}{21}{16.17}{13.50}{22}
\emline{17.50}{14.17}{23}{19.33}{14.33}{24}
\emline{20.50}{14.50}{25}{21.67}{14.50}{26}
\emline{22.67}{14.50}{27}{24.00}{14.33}{28}
\emline{5.50}{4.50}{29}{7.67}{4.50}{30}
\emline{9.50}{4.50}{31}{11.50}{4.50}{32}
\emline{13.00}{4.50}{33}{15.50}{4.50}{34}
\emline{17.50}{4.50}{35}{19.83}{4.50}{36}
\emline{21.00}{4.50}{37}{23.50}{4.50}{38}
\end{picture}
\end{array}
&
\begin{array}{c}
\begin{picture}(20.00,20.00)
\put(10.00,10.00){\circle*{0.93}}
\emline{9.33}{10.65}{1}{0.08}{20.00}{2}
\emline{9.33}{9.14}{3}{0.00}{0.00}{4}
\emline{11.08}{9.25}{5}{20.00}{0.00}{6}
\emline{11.00}{10.65}{7}{20.00}{20.00}{8}
\emline{3.00}{18.00}{9}{4.00}{17.96}{10}
\emline{5.00}{18.00}{11}{6.00}{17.96}{12}
\emline{7.00}{18.00}{13}{8.00}{17.96}{14}
\emline{9.00}{18.00}{15}{10.00}{17.96}{16}
\emline{11.00}{18.00}{17}{12.00}{17.96}{18}
\emline{13.00}{18.00}{19}{14.00}{17.96}{20}
\emline{15.00}{18.00}{21}{16.00}{17.96}{22}
\emline{17.00}{18.00}{23}{18.00}{17.96}{24}
\end{picture}
\end{array}
&
\begin{array}{c}
\begin{picture}(20.00,20.00)
\put(10.00,10.00){\circle*{0.93}}
\emline{9.33}{10.65}{1}{0.08}{20.00}{2}
\emline{9.33}{9.14}{3}{0.00}{0.00}{4}
\emline{11.08}{9.25}{5}{20.00}{0.00}{6}
\emline{11.00}{10.65}{7}{20.00}{20.00}{8}
\emline{18.00}{17.00}{9}{18.00}{16.02}{10}
\emline{18.00}{15.00}{11}{18.00}{13.98}{12}
\emline{18.00}{13.00}{13}{18.00}{12.04}{14}
\emline{18.00}{11.00}{15}{18.00}{10.00}{16}
\emline{18.00}{9.00}{17}{18.00}{7.96}{18}
\emline{18.00}{7.00}{19}{18.00}{6.02}{20}
\emline{18.00}{5.00}{21}{18.00}{3.98}{22}
\emline{18.00}{3.00}{23}{18.00}{2.04}{24}
\end{picture}
\end{array}
\\
k=&0&1&2\\
g_k&\
\begin{array}{c}
\begin{picture}(30.00,20.00)
\put(20.00,10.00){\circle*{0.93}}
\emline{9.33}{10.65}{1}{0.08}{20.00}{2}
\emline{9.33}{9.14}{3}{0.00}{0.00}{4}
\emline{21.08}{9.25}{5}{30.00}{0.00}{6}
\emline{21.00}{10.65}{7}{30.00}{20.00}{8}
\put(10.00,10.00){\circle*{0.93}}
\put(11.00,9.67){\rule{7.50\unitlength}{0.50\unitlength}}
\emline{7.00}{5.83}{9}{9.33}{5.83}{10}
\emline{9.33}{5.83}{11}{9.33}{5.83}{12}
\emline{10.67}{5.83}{13}{13.00}{6.17}{14}
\emline{14.33}{6.50}{15}{16.00}{7.17}{16}
\emline{16.67}{7.67}{17}{17.33}{9.00}{18}
\emline{12.33}{10.83}{19}{13.50}{12.17}{20}
\emline{14.33}{12.83}{21}{16.17}{13.50}{22}
\emline{17.50}{14.17}{23}{19.33}{14.33}{24}
\emline{20.50}{14.50}{25}{21.67}{14.50}{26}
\emline{22.67}{14.50}{27}{24.00}{14.33}{28}
\emline{5.50}{4.50}{29}{7.67}{4.50}{30}
\emline{9.50}{4.50}{31}{11.50}{4.50}{32}
\emline{13.00}{4.50}{33}{15.50}{4.50}{34}
\emline{17.50}{4.50}{35}{19.83}{4.50}{36}
\emline{21.00}{4.50}{37}{23.50}{4.50}{38}
\end{picture}
\end{array}
&
\begin{array}{c}
\begin{picture}(30.00,20.00)
\put(20.00,10.00){\circle*{0.93}}
\emline{9.33}{10.65}{1}{0.08}{20.00}{2}
\emline{9.33}{9.14}{3}{0.00}{0.00}{4}
\emline{21.08}{9.25}{5}{30.00}{0.00}{6}
\emline{21.00}{10.65}{7}{30.00}{20.00}{8}
\put(10.00,10.00){\circle*{0.93}}
\put(11.00,9.67){\rule{7.50\unitlength}{0.50\unitlength}}
\emline{7.00}{5.83}{9}{9.33}{5.83}{10}
\emline{9.33}{5.83}{11}{9.33}{5.83}{12}
\emline{10.67}{5.83}{13}{13.00}{6.17}{14}
\emline{14.33}{6.50}{15}{16.00}{7.17}{16}
\emline{16.67}{7.67}{17}{17.33}{9.00}{18}
\emline{12.33}{10.83}{19}{13.50}{12.17}{20}
\emline{14.33}{12.83}{21}{16.17}{13.50}{22}
\emline{17.50}{14.17}{23}{19.33}{14.33}{24}
\emline{20.50}{14.50}{25}{21.67}{14.50}{26}
\emline{22.67}{14.50}{27}{24.00}{14.33}{28}
\emline{5.50}{4.50}{29}{7.67}{4.50}{30}
\emline{9.50}{4.50}{31}{11.50}{4.50}{32}
\emline{13.00}{4.50}{33}{15.50}{4.50}{34}
\emline{17.50}{4.50}{35}{19.83}{4.50}{36}
\emline{21.00}{4.50}{37}{23.50}{4.50}{38}
\emline{6.33}{15.33}{39}{7.17}{15.33}{40}
\emline{9.00}{15.33}{41}{10.67}{15.33}{42}
\emline{12.50}{15.33}{43}{14.33}{15.33}{44}
\emline{16.17}{15.33}{45}{18.33}{15.33}{46}
\emline{20.17}{15.33}{47}{22.00}{15.33}{48}
\emline{23.67}{15.33}{49}{25.00}{15.33}{50}
\end{picture}
\end{array}
&
\begin{array}{c}
\begin{picture}(30.00,20.00)
\put(20.00,10.00){\circle*{0.93}}
\emline{9.33}{10.65}{1}{0.08}{20.00}{2}
\emline{9.33}{9.14}{3}{0.00}{0.00}{4}
\emline{21.08}{9.25}{5}{30.00}{0.00}{6}
\emline{21.00}{10.65}{7}{30.00}{20.00}{8}
\put(10.00,10.00){\circle*{0.93}}
\put(11.00,9.67){\rule{7.50\unitlength}{0.50\unitlength}}
\emline{7.00}{5.83}{9}{9.33}{5.83}{10}
\emline{9.33}{5.83}{11}{9.33}{5.83}{12}
\emline{10.67}{5.83}{13}{13.00}{6.17}{14}
\emline{14.33}{6.50}{15}{16.00}{7.17}{16}
\emline{16.67}{7.67}{17}{17.33}{9.00}{18}
\emline{12.33}{10.83}{19}{13.50}{12.17}{20}
\emline{14.33}{12.83}{21}{16.17}{13.50}{22}
\emline{17.50}{14.17}{23}{19.33}{14.33}{24}
\emline{20.50}{14.50}{25}{21.67}{14.50}{26}
\emline{22.67}{14.50}{27}{24.00}{14.33}{28}
\emline{5.50}{4.50}{29}{7.67}{4.50}{30}
\emline{9.50}{4.50}{31}{11.50}{4.50}{32}
\emline{13.00}{4.50}{33}{15.50}{4.50}{34}
\emline{17.50}{4.50}{35}{19.83}{4.50}{36}
\emline{21.00}{4.50}{37}{23.50}{4.50}{38}
\emline{6.33}{15.33}{39}{7.17}{15.33}{40}
\emline{9.00}{15.33}{41}{10.67}{15.33}{42}
\emline{12.50}{15.33}{43}{14.33}{15.33}{44}
\emline{16.17}{15.33}{45}{18.33}{15.33}{46}
\emline{20.17}{15.33}{47}{22.00}{15.33}{48}
\emline{23.67}{15.33}{49}{25.00}{15.33}{50}
\emline{27.83}{17.17}{51}{27.83}{15.17}{52}
\emline{27.83}{13.83}{53}{27.83}{11.83}{54}
\emline{27.83}{10.17}{55}{27.83}{8.17}{56}
\emline{27.83}{6.50}{57}{27.83}{4.83}{58}
\emline{27.83}{4.83}{59}{27.83}{4.50}{60}
\emline{27.83}{3.33}{61}{27.83}{2.50}{62}
\end{picture}
\end{array}
\end{array}
\]
\caption{\label{figIIa}
Correspondence of a family of graphs $(g_0, g_1^0, g_2^0)$ to a nest
of subgraphs $g_0 \subset g_1 \subset g_2 $.}
\end{figure}
\end{document}